\documentclass[twocolumn]{aastex6} 
\usepackage{graphicx}
\usepackage{epstopdf}
\usepackage{natbib}
\usepackage{rotating}
\usepackage{makecell}

\def\be{\begin{equation}}
\def\ee{\end{equation}}
\def\bea{\begin{eqnarray}}
\def\eea{\end{eqnarray}}
 
\newcommand{\HI}{\hbox{{\rm H}\kern 0.1em{\sc i}}}
\newcommand{\HII}{\hbox{{\rm H}\kern 0.1em{\sc ii}}}
\newcommand{\NII}{\hbox{{\rm N}\kern 0.1em{\sc ii}}}

\begin{document}

\title{CHANG-ES X: Spatially-resolved Separation of Thermal Contribution from Radio Continuum Emission in Edge-on Galaxies}

\author{Carlos J. Vargas\altaffilmark{1}, Silvia Carolina Mora-Partiarroyo\altaffilmark{2}, Philip Schmidt\altaffilmark{2}, Richard J. Rand\altaffilmark{3}, Yelena Stein\altaffilmark{4}, Ren\'{e} A. M.  Walterbos\altaffilmark{1}, Q. Daniel Wang\altaffilmark{5}, Aritra Basu\altaffilmark{2},  Maria Patterson\altaffilmark{6}, Amanda Kepley\altaffilmark{7}, Rainer Beck\altaffilmark{2}, Judith Irwin\altaffilmark{8}, George Heald\altaffilmark{9}, Jiangtao Li\altaffilmark{10}, Theresa Wiegert\altaffilmark{8}}

\altaffiltext{1}{Department of Astronomy, New Mexico State University, Las Cruces, NM 88001, U.S.A.}
\altaffiltext{2}{Max-Planck-Institut f{\"u}r Radioastronomie, Auf dem H{\"u}gel 69, 53121 Bonn, Germany}
\altaffiltext{3}{Department of Physics and Astronomy, University of New Mexico, 1919 Lomas Blvd. NE, Albuquerque, NM 87131, U.S.A.}
\altaffiltext{4}{Astronomisches Institut, Ruhr-Universit{\"a}̈t Bochum, Universit{\"a}tsstr. 150, 44780 Bochum, Germany}
\altaffiltext{5}{Department of Astronomy, University of Massachusetts, 710 North Pleasant St., Amherst, MA 01003, U.S.A.}
\altaffiltext{6}{Department of Astronomy, University of Washington, Box 351580, Seattle, WA 98195, U.S.A.}
\altaffiltext{7}{NRAO  Charlottesville, 520  Edgemont  Road,  Charlottesville,  VA, 22903-2475, U.S.A.}
\altaffiltext{8}{Queen's University, Kingston, ON, Canada}
\altaffiltext{9}{CSIRO Astronomy and Space Science, 26 Dick Perry Ave, Kensington WA 6151, Australia}
\altaffiltext{10}{Dept.  of Astronomy, University of Michigan, 311 West Hall, 1085 S. University Ave., Ann Arbor, MI 48109, U.S.A.}
\begin{abstract}

We analyze the application of star formation rate (SFR) calibrations using H$\alpha$ and $22$ micron infrared imaging data in predicting the thermal radio component for a test sample of 3 edge-on galaxies (NGC 891, NGC 3044, and NGC 4631) in the Continuum Halos in Nearby Galaxies -- an EVLA Survey (CHANG-ES). We use a mixture of H$\alpha$ and $24$ micron calibration from \citet{calzettietal07}, and a linear $22$ micron only calibration from \cite{jarrettetal13} on the test sample. We apply these relations on a pixel-to-pixel basis to create thermal prediction maps in the two CHANG-ES bands: L- and C-band (1.5 GHz and 6.0 GHz, respectively). We analyze the resulting non-thermal spectral index maps, and find a characteristic steepening of the non-thermal spectral index with vertical distance from the disk after application of all methods. We find possible evidence of extinction in the $22$ micron data as compared to $70$ micron Spitzer Multband Imaging Photometer (MIPS) imaging in NGC 891. We analyze a larger sample of edge-on and face-on galaxy $25$ micron to $100$ micron flux ratios, and find that the ratios for edge-ons are systematically lower by a factor of $1.36$, a result we attribute to excess extinction in the mid-IR in edge-ons. We introduce a new calibration for correcting the H$\alpha$ luminosity for dust when galaxies are edge-on or very dusty. 

\end{abstract}

\section{Introduction}

Radio continuum emission from active star-forming galaxies contain both free-free (thermal) and synchrotron (non-thermal) emission. Separating the emission from each component allows for each to be studied individually. Under the assumption of equipartition, analyses of the non-thermal component allows for characterization of magnetic fields and the cosmic rays (CRs) that illuminate those fields. 

In cases where data from two discrete radio continuum frequencies are available, separation of the thermal and non-thermal components has been carried out by assuming a constant non-thermal spectral index, e.g. \citet{kleinetal82}. One of the first papers with an advanced separation of thermal and nonthermal radio emission other than assuming a constant nonthermal spectral index is that by \citet{becketal82}, where the thermal emission from M31 was estimated from a catalog of HII regions, corrected for extinction. The bulk of cosmic ray particles are believed to be accelerated in supernova shocks in star-forming regions. As CRs propagate outward, they suffer energy losses which result in a steepening of the synchrotron spectral index \citep{condon92}. Since assuming a constant non-thermal spectral index makes it impossible to study variations due to CR production and energy losses in the spectral index distribution, an alternate separation approach must be employed. 

Both free-free emission and recombination line emission originate from ionized regions, making recombination lines a good tracer for thermal emission. H$\alpha$, in particular, is the most preferred recombination line to observe, since it is the strongest Balmer line. However, H$\alpha$ emission is strongly obscured by dust along the line of sight and needs to be corrected for extinction, especially in edge-on galaxies. It is also worth noting that radio recombination lines are not subject to extinction along the line of sight, and would thus be ideal for tracing thermal emission. However, these lines are too faint to map on a resolved basis with current technology from the diffuse ionized regions in nearby galaxies. \citet{tabatabaeietal07} used Effelsberg $6.2$ cm observations in an attempt to map radio recombination line emission in M33, but were unsuccessful. We note that extragalactic carbon radio recombination lines have been detected in M82 with LOFAR \citep{morabitoetal14}. The authors of that study infer that the detected carbon lines are likely associated with cold atomic gas near the nucleus of M82.

Dust grains emitting at IR wavelengths can be heated by ionizing and non-ionizing photons from both young and evolved stars, as well as AGN. Empirical relations have been established relating mid-IR emission to star formation, and thus thermal emission from {\HII} regions. These empirical relations rely on the assumption that the distribution of dust heating mechanisms other than young stars (ie. evolved stars, AGN, etc.) are negligible, or similar to the distribution of young stars. Also, mid-IR emission originates from substantially heated grains, thus the correlation with young, massive stars is likely robust. Additionally, H$\alpha$ emission and $24$ micron emission show spatial correlation in face-on galaxies, thus suggesting that dust shells around evolved stars is not a significant contribution to mid-IR emission. AGN contribution likely skews these empirical relations in the central regions of galaxies which host AGN. However, separate observations tailored to search for AGN are needed to determine if a galaxy is a host. 

The emission from heated dust grains can be empirically related to the extinction of H$\alpha$ emission, allowing extinction-corrected H$\alpha$ emission to be used as a SFR tracer. Star formation in both obscured and unobscured regions can be traced by a combination of H$\alpha$ and $24$ micron measurements \citep{kennicuttetal07}. \citet{calzettietal07} found that H$\alpha$ emission corrected using $24$ micron data adequately traces extinction-corrected Pa$\alpha$ (a star formation tracer less affected by extinction than H$\alpha$) emission, in the {\HII} regions of 33 galaxies in the Spitzer Infrared Nearby Galaxies Survey (SINGS). Furthermore, \citet{kennicuttetal09} found good agreement between integrated H$\alpha$ emission corrected for extinction with mixed H$\alpha$ and $24$ micron luminosities. They also find good agreement using H$\alpha$/H$\beta$ ratios in optical spectra, where deviations from a theoretical constant ratio in the ISM allow for an H$\alpha$ extinction correction.

To date, several studies have estimated thermal emission in spiral galaxies at various inclinations, i.e. \citet{tabatabaeietal07}, \citet{basuetal12}, \citet{leroyetal12}, \citet{basuetal17}. However, galaxies viewed edge-on add additional complication. Line-of-sight H$\alpha$ extinction is certainly larger in the midplanes of edge-on galaxies, since photons are more likely to encounter dust particles on their long paths. Thus, the observed H$\alpha$ may be only from sources residing on the near side of the galaxy. Additionally, structures like spiral arms, bars, and distortions make spiral galaxies non-homogeneous. Hence, the observed H$\alpha$ emission may not be representative of the full disk emission. Dust extinction effects decrease with increasing wavelength, and so IR emission should be largely unaffected.

The Continuum Halos in Nearby Galaxies -- an EVLA Survey (CHANG-ES; \citealt{irwinetal12, irwinetal12b, irwinetal13, wiegertetal15, irwinetal15, lietal16, damasetal16, irwinetal16}) aims to establish the connection between star formation in galaxy disks and non-thermal processes, like halo magnetic fields, CR injection and propagation, and active galactic nuclei (AGN). To accomplish this goal, CHANG-ES has observed 35 highly-inclined galaxies in the nearby universe with the NSF's Karl G. Jansky Very Large Array (VLA) in B, C, and D configurations. The variety of VLA antenna configurations allow for study on various spatial scales. The observations were performed in L-band (centered at $1.5$~GHz) and C-band (centered at $6$~GHz) in all polarization products. CHANG-ES improves on past radio continuum surveys in the large sample size, diversity of physical sizes within the sample, and the large bandwidths obtained with the Wideband Interferometric Digital ARchitecture (WIDAR) correlator at the VLA. The latter is perhaps the most drastic improvement: WIDAR allows the CHANG-ES observations to be, in some cases, an order of magnitude more sensitive to continuum emission than previous studies, and also allows for an improved study of the regular magnetic field morphology by applying the Rotation Measure Synthesis algorithm \citep{burn66,brentjensetal05}.

\citet{lietal16} found that the radio-IR correlation is shifted for CHANG-ES galaxies from those found for face-on galaxies. The shift implied that the mid-IR emission could be subject to extinction, itself. Furthermore, the mid-IR SFR estimates used by that study, and calculated in \citet{wiegertetal15} for the CHANG-ES sample, were a factor of $\sim2.3$ lower than those from IRAS total infrared (TIR) estimates in \cite{irwinetal12}. While differing SFR calibrations for TIR and mid-IR emission may explain some of this discrepancy, it may also indicate that $22$ micron emission may suffer from measurable extinction in edge-ons, and may need correction prior to SFR estimation and thermal prediction. 

Due to the added complexity of an edge-on perspective, there is no standard method for estimating the thermal component of edge-on galaxies, especially at small spatial scales. Hence, this study will examine the validity of predicting the thermal emission in edge-on galaxies using both ancillary H$\alpha$ and mid-IR imaging as a proxy for thermal emission in a subsample of $3$ galaxies from the CHANG-ES survey: NGC~891, NGC~3044, and NGC~4631, and aim to identify a consistent method of thermal prediction for edge-on galaxies. Calibrated radio continuum images and methodology from Ph.D. theses \citet{schmidt} and \citet{mora} are used in this study for NGC~891 and NGC~4631, respectively. 


\section{Data}
\label{datasection}

We identified three galaxies within the CHANG-ES survey with pre-existing and readily available H$\alpha$ imaging of their entire disks, in order to test methods of independently estimating the thermal radio component using both H$\alpha$ and Wide-field Infrared Survey Explorer (WISE; \citealt{wrightetal10}) $22$ micron mid-infrared imaging. The three galaxies chosen for this test sample were NGC 891, NGC 3044, and NGC 4631. In addition to the current availability of H$\alpha$ imaging for these galaxies, NGC 891 and NGC 4631 are both well-studied. NGC 3044 has a relatively high surface SFR, and is smaller in size than the other two, making a short spacings correction less of a concern. Properties of each are listed in Table \ref{sampletable}.

\begin{table*}
\center
\begin{tabular}{ c c c c c c c  } 

 \hline
 \hline

Galaxy & Type & D (Mpc) & $d_{25}$ $(\arcmin)$ & $d_{25}$ (kpc) & \thead{SFR \\ (M$_{\odot}/$yr)} & \thead{Surface SFR \\ ($\times 10^{-3}$ M$_{\odot}/$yr$/$kpc$^2$)} \\
 \hline
 NGC 891 &  Sb   & 9.1 & 12.2 & 33.6 & 1.55 & 3.13    \\
 NGC 3044 & SBc  & 20.3 & 4.4  & 26.0 & 0.95 & 3.70   \\
 NGC 4631 & SBcd & 7.4 & 14.7 & 32.3 & 1.33 & 3.10   \\
 \hline
 \hline
\end{tabular}

\caption{Properties of test sample galaxies. The total SFR and surface SFR are calculated from the WISE $22$ micron imaging and are quoted for the entire CHANG-ES sample in \cite{wiegertetal15}.}
\label{sampletable}

\end{table*}

\begin{figure*}
\centering
\includegraphics[scale=0.65]{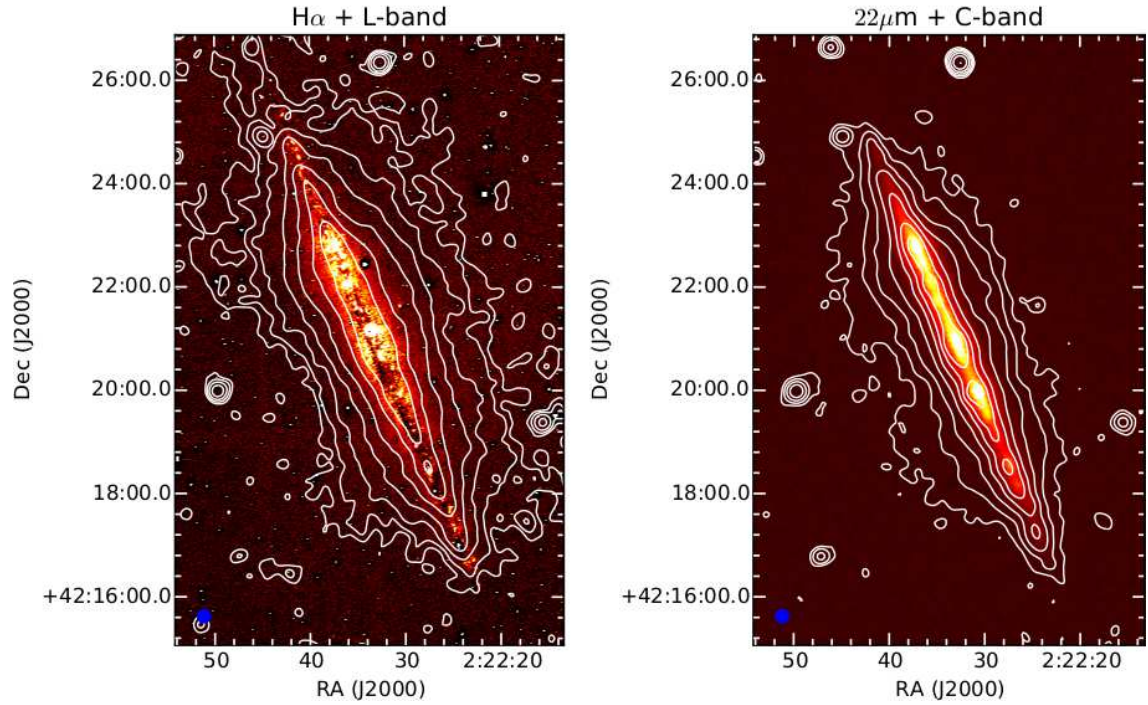}
\caption{\small Input data for NGC 891. Left: H$\alpha$ with L-band contours overlaid; Right: WISE $22$ micron with C-band contours overlaid. All images are shown at their highest resolution (H$\alpha$ - $1\arcsec$, $22$ micron - $12\arcsec$). Contours on the L- band and C- band images begin at $3\sigma$ and increase in multiples of $2$. The $3\sigma$ lowest contour for the L-band image is at $75~\mu$Jy beam$^{-1}$, and that of the C-band image is $24~\mu$Jy beam$^{-1}$. The beam size for the C-band contours is $9.0\arcsec~\times~8.8\arcsec$, and that of the L-band contours is $10.5\arcsec~\times~9.7\arcsec$. The beams of the overlaid radio images are shown in the lower left of each respective panel. Both the H$\alpha$ and $22$ micron images are shown in a logarithmic stretch.} 

\label{891input}
\end{figure*}

\begin{figure*}
\centering
\includegraphics[scale=0.39]{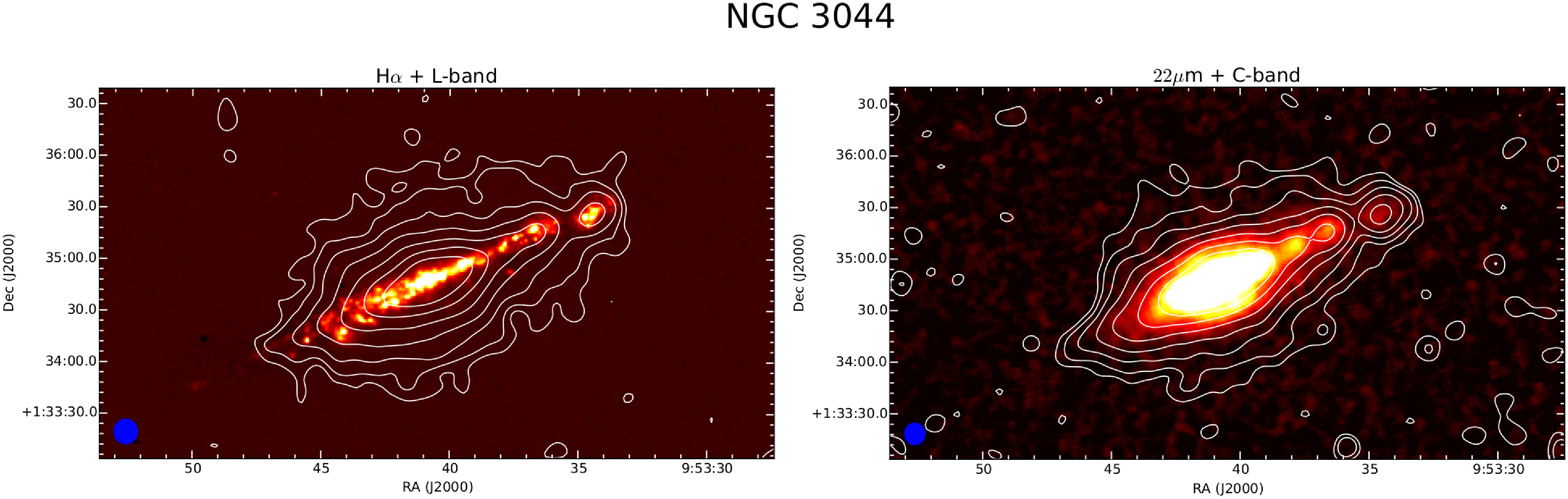}
\caption{\small Input data for NGC 3044. Left: H$\alpha$ with L-band contours overlaid; Right: WISE $22$ micron with C-band contours overlaid. Contours on the L- band and C- band images begin at $3\sigma$ and increase in multiples of $2$. The $3\sigma$ lowest contour for the L-band image is at $63~\mu$Jy beam$^{-1}$, and that of the C-band image is $14~\mu$Jy beam$^{-1}$. The beam size for the C-band contours is $10.7\arcsec~\times~9.6\arcsec$, and that of the L-band contours is $11.4\arcsec~\times~10.6\arcsec$. The beams of the overlaid radio images are shown in the lower left of each respective panel. H$\alpha$ and $22$ micron image resolutions are the same as that of Figure \ref{891input}, and are shown in a logarithmic stretch.  }
\label{3044input}
\end{figure*}

\begin{figure*}
\centering
\includegraphics[scale=0.55]{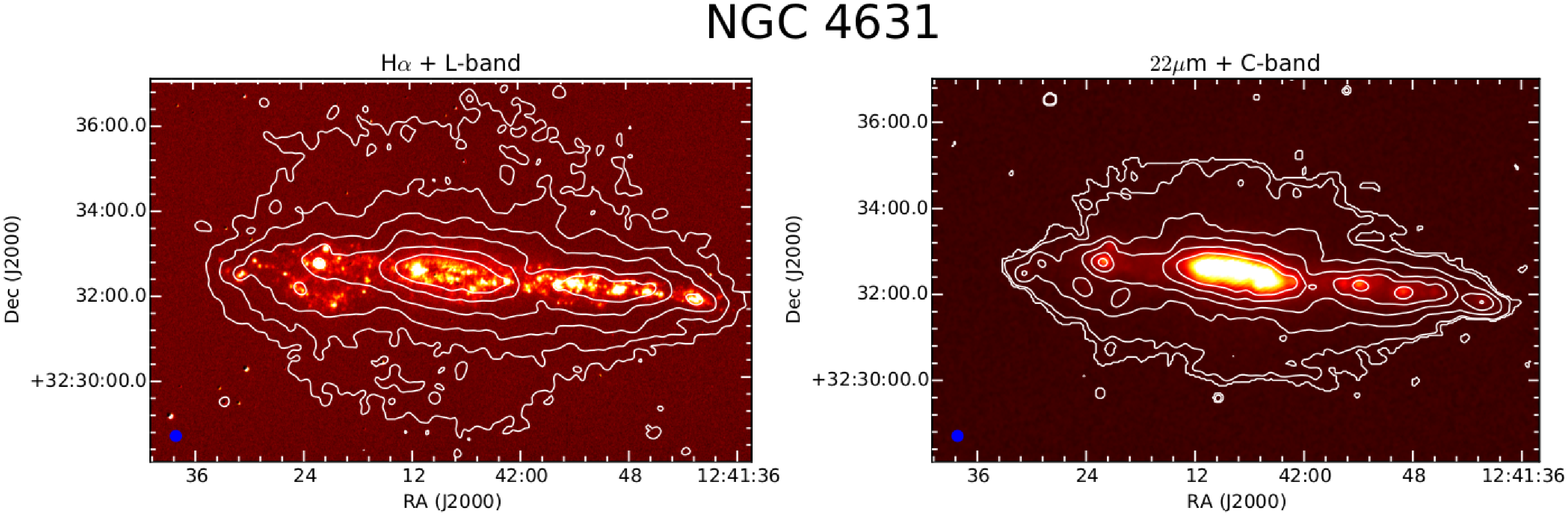}
\caption{\small Input data for NGC 4631. Left: H$\alpha$ with L-band contours overlaid; Right: WISE $22$ micron with C-band contours overlaid. Contours on the L- band and C- band images begin at $3\sigma$ and increase in multiples of $2$. The $3\sigma$ lowest contour for the L-band image is at $70~\mu$Jy beam$^{-1}$, and that of the C-band image is $21~\mu$Jy beam$^{-1}$. The beam size for the C-band contours is $8.9\arcsec~\times~8.6\arcsec$, and that of the L-band contours is $10.1\arcsec~\times~9.9\arcsec$. The beams of the overlaid radio images are shown in the lower left of each respective panel. H$\alpha$ and $22$ micron image resolutions are the same as that of Figure \ref{891input}, and are shown in a logarithmic stretch. }
\label{4631input}
\end{figure*}

We present the H$\alpha$, WISE $22$ micron, and CHANG-ES radio input data in Figures \ref{891input}, \ref{3044input}, and \ref{4631input} for the test sample of NGC 891, NGC 3044, and NGC 4631, respectively, and describe each data set in the remainder of this section.

\subsection{CHANG-ES Data}

CHANG-ES observations in C-band were centered at $6$~GHz and have a bandwidth of $2$~GHz in $16$ spectral windows, each divided into $64$ channels. In L-band, observations were centered at $1.5$~GHz  and have a bandwidth of $512$~MHz in $32$ spectral windows, each divided into $2048$ channels.

CHANG-ES data taken in VLA C and D configurations at both frequency bands were used.  See Table 2 in \citet{irwinetal12} for the theoretical noise limits and resolutions of the various observations, and refer to Table 4 and 5 in \citet{wiegertetal15} for the resulting rms and beam size of the D array observations. All radio maps used for this analysis are primary beam corrected.

The radio data used for NGC 891 and NGC 4631 are corrected for short spacings using single-dish Effelsberg 100-m telescope observations, taken with the purpose of supplementing the CHANG-ES data. The optical diameter of NGC 3044 is $\sim4.4\arcmin$. The largest detectable sizes of the CHANG-ES observations are $16\arcmin$ at L-band, and $4\arcmin$ at C-band (see \citealt{irwinetal12}). Therefore, missing flux at L-band for NGC 3044 will be negligible, and missing flux at C-band will be low. For details on the Effelsberg observations and the combination of single-dish and interferometer data see \citet{schmidt}, and \citet{mora}. 

\subsection{H$\alpha$ Imaging}

The H$\alpha$ imaging data for NGC 891 was taken at Kitt Peak National Observatory (KPNO) using the 4-m telescope, by author MP. The narrowband H$\alpha$ filter used has a central wavelength of $6574.74 \rm{\AA}$ and a full width at half maximum (FWHM) of $80.62\rm{\AA}$, and was chosen to encompass the full velocity range of the galaxy. Narrowband frames were taken in the format of $3\times 15$ minute exposures, and corresponding R band images were taken. Images are bias subtracted, crosstalk removed, flat fielded with dome flats, sky subtracted to level zero, and finally an R band image is scaled and subtracted from the narrowband image. The H$\alpha$ image for NGC 891 is corrected for [{\NII}] contamination assuming a ratio of [{\NII}]/H$\alpha$ $=0.4$. The rms noise of the resulting continuum subtracted H$\alpha$ image is $7.1\times10^{-18}$ erg cm$^{-2}$ s$^{-1}$ arcsec$^{-2}$ (EM $=~3.45$ pc cm$^{-6}$, assuming $10^4$ K).

For NGC 3044, we use the H$\alpha$ imaging presented in \citet{collinsetal00}. This image was obtained in $4$ hours of integration time at the KPNO using the 0.9-m telescope. The narrowband filter used had a width of $28 \rm{\AA}$ and was chosen to encompass the redshifted H$\alpha$ line. Continuum images were taken before and after narrowband exposures so the stellar continuum could be subtracted. The H$\alpha$ image of NGC 3044 contains very little [{\NII}] contamination due to the throughput of the [{\NII}] lines within the filter used, and so no further correction is used. The rms noise of the continuum subtracted H$\alpha$ image is $1.1\times10^{-17}$ erg cm$^{-2}$ s$^{-1}$ arcsec$^{-2}$ (EM $=~5.31$ pc cm$^{-6}$, assuming $10^4$ K).

Continuum subtracted H$\alpha$ data for NGC 4631 were available from the Spitzer Infrared Nearby Galaxy Survey (SINGS; \citealt{kennicuttetal03}). Observations were done at KPNO using the KP1563 filter (width $67~\rm{\AA}$) in 2002. Narrowband H$\alpha$ observations were taken in two $15$ minute exposures for cosmic ray removal. The original data also contain [{\NII}] emission from the doublet at $\lambda$ $6548$, $6583\mathrm{\AA}$, but we apply a corrected for [{\NII}] contamination assuming a ratio of [{\NII}]/H$\alpha$ $=0.4$. The original map has a pixel scale of $0.305\arcsec$. The rms noise of the continuum subtracted H$\alpha$ image is $1.9\times10^{-17}$ erg cm$^{-2}$ s$^{-1}$ arcsec$^{-2}$ (EM $=~9.31$ pc cm$^{-6}$, assuming $10^4$ K).

\subsection{WISE $22$ Micron Imaging}

Wide Field Infrared Explorer (WISE; \citealt{wrightetal10}) $22$ micron imaging were kindly provided by T. Jarrett with enhanced resolution of $12\arcsec.4$, using the WISE Enhanced Galaxy Resolution Atlas (WERGA) process \citep{jarrettetal12} for every CHANG-ES galaxy. The WERGA pipeline used point spread function (PSF) subtraction to remove foreground stars, and a constant background was subtracted. An aperture correction for extended sources, a color correction, and calibration correction for spiral galaxies were applied as in \citet{jarrettetal13}. No corrections were made for extinction. In what follows, when $24$ micron fluxes are needed, we obtain them from $22$ micron fluxes by multiplying by a factor of $1.03$, as found from the tight linear relation of the two fluxes by \citet{wiegertetal15} for CHANG-ES galaxies (see also \citealt{jarrettetal13}). The maps were convolved to a Gaussian beam with a FWHM of $15\arcsec$ using the kernels provided by \citet{anianoetal11}, as those kernels are the closest to the WISE WERGA data resolution, without exceeding its native resolution (note: this is not necessary for the H$\alpha$ imaging, which all have Gaussian PSFs). After convolution, the images were regridded to the geometry of the radio maps.

\section{Methodology}
\label{methodology}
The main goal of this paper is to create and analyze maps of the predicted thermal radio continuum emission (thermal prediction maps, hereafter) for three edge-on galaxies using conventional SFR tracers and calibrations, and to then identify a general method of thermal prediction for edge-on galaxies. To that end, we analyze relations between H$\alpha$ and $24$ micron emission and the extinction-corrected H$\alpha$ emission. 

We begin with a mid-infrared monochromatic relation to SFR from \citet{jarrettetal13}:
\begin{equation}
\label{SFRequation22}
\rm{SFR}_{22\mu\rm{m}}(\pm 0.04)(M_{\odot} yr^{-1})=7.50(\pm 0.07) \times 10^{-10}\nu L_{22}(L_{\odot})
\end{equation}

This is a direct relation between $22$ micron (rather than $24$ micron) intensity and SFR, which is directly related to thermal flux, and also has the added benefit of being linear. We also use this method for consistency with CHANG-ES Paper IV \citep{wiegertetal15}, in which this SFR calibration was used. We refer to this method as the `$22$ micron Only Method', hereafter. We note the existence of other mid-infrared monochromatic relations to SFR or extinction-corrected H$\alpha$ emission, such as \citet{relanoetal07}. These calibrations using the $24$ micron band are non-linear, and thus need to be linearized for use on a pixel-to-pixel basis, else the sums of the pixels in the resulting image will be a factor of $\sim3$ larger than the predicted sum from integrated values.  We effectively linearized the calibration from \citet{relanoetal07} by forcing the sum of each thermal prediction map derived using this method to be the same as the sum using integrated values. The average relative difference between the predicted thermal flux results using the linearized \citet{relanoetal07} relation, and the relation from \citet{jarrettetal13} was $\sim 5\%$. 
In light of their similarity, we elected to move forward using the linear $22$ micron calibration from \citet{jarrettetal13}.

We also use a combination of H$\alpha$ and $24$ micron emission from \citet{calzettietal07}, which we refer to as the `mixture method', hereafter:
\begin{equation}
\label{Halpha+24microns}
\rm{L(H\alpha_{corr}}) = L(\rm{H\alpha_{obs}}) + a \cdot \nu L_{\nu}(24 \rm{\mu m})
\end{equation}
The coefficient $a$ is effectively a weighting factor for the $24$ micron contribution to the mixture. We adopt a value of $a=0.031$, which was empirically measured by \citet{calzettietal07} with a stated uncertainty of $0.006$ ($\sim20\%$). We elect to use $a=0.031$ instead of $a=0.02$ from \citet{kennicuttetal09} since the former was estimated in {\HII} regions, rather than integrated galaxies. The individual pixels for which we calculate the SFR are closer to the size scale of {\HII} regions than to that of entire galaxies. More specifically, the pixel size in the final versions of each map is $2.5\arcsec$, which corresponds to physical sizes of $110$ pc, $246$ pc, and $90$ pc at the assumed distances of NGC 891, NGC 3044, and NGC 4631, respectively. The relations we analyze assume {\HII} regions sizes of $100-200$ pc. 

We note all empirical relations to SFR used in this work assume the same IMF.  

We then relate the extinction corrected H$\alpha$ emission to SFR using the following relation from \citet{murphyetal11}:

\begin{equation}
\label{SFRequation}
\left( \frac{\rm{SFR}_{\rm{mix}}}{\rm{M_{\odot}}\rm{yr}^{-1}} \right) = 5.37\times 10^{-42} \left( \frac{\rm{L(\rm{H\alpha_{obs}})} +a\cdot \nu \rm{L}_{\nu}(24 \mu \rm{m})}{\rm{erg} \cdot \rm{s}^{-1}}\right)
\end{equation}

Finally, for both methods, assuming case B recombination, we calculate the thermal radio component in L- and C- bands directly from the SFR found from Equation \ref{SFRequation22} or \ref{SFRequation}, by combining with Equation 11 from \citet{murphyetal11}:
\begin{equation}
\label{radioequation}
\left(\frac{\rm{L}^{\rm{T}}_{\nu}}{\rm{erg} \cdot \rm{s}^{-1} \rm{Hz}^{-1}} \right)=2.2\times 10^{27} \left(\frac{\rm{T_e}}{10^4 \rm{K}}\right)^{0.45} \left(\frac{\nu}{\rm{GHz}}\right)^{-0.1} \left(\frac{\rm{SFR}^{T}_{\nu}}{\rm{M_{\odot} {yr}^{-1}}}\right)
\end{equation}

We assume an electron temperature of $\rm{T_e} = 10,000$ K, and consider uncertainties in this key parameter in the following subsection. The estimates of SFR from the thermal indicators, SFR$_{22\mu\rm{m}}$ and SFR$_{\rm{mix}}$, are used in Equation \ref{radioequation} as SFR$_{\nu}^{\rm{T}}$. 

We note the existence of another possible method of estimating thermal emission in galaxies, used in \cite{tabatabaeietal07}, where dust optical depth maps are generated by fitting the FIR SED. However, this method must assume the shape of the radiation field, which would likely change dramatically along various lines of sight in the edge-on case. This method also must assume a geometry such that the dust surface density maps may be applied to the H$\alpha$ data at each pixel to correct for extinction. In the edge-on case, there are likely some regions where H$\alpha$ emission suffers from so much extinction that it would not be detected. In such regions, the application of this method is unclear. Lastly, we lack the data needed to reproduce this analysis for the CHANG-ES sample.   

\subsection{Uncertainties}

For both the input H$\alpha$ and $22$ micron imaging, we estimate background variations that contribute to uncertainties. We estimate the background uncertainty by finding the root-mean-squared value in ten $100\times100$ pixel boxes in regions with no galaxy emission. We quantify the uncertainty per pixel as the mean of these ten root-mean-squared background values. 

We estimate the uncertainties in [{\NII}] emission line subtraction within the H$\alpha$ imaging by estimating the line's throughput in the used narrowband H$\alpha$ filters. According to \citet{collinsetal00}, the throughput of both [{\NII}] lines in NGC 3044 is very low. Thus, we conservatively add an uncertainty of $10\%$ to the H$\alpha$ imaging of that galaxy to account for potential [{\NII}] contamination. The H$\alpha$ images for NGC 891 and NGC 4631 both contain emission from the [{\NII}] line within their narrowband filter, and the line contribution is accounted for (see Section \ref{datasection}). We add an uncertainty to the treatment of the [{\NII}] line flux of $10\%$ to H$\alpha$ measured fluxes.

The electron temperature $\rm{T_e}$ is also an uncertain quantity. We assume the electron temperature to range between $7,000 - 13,000$ K (ie. $\rm{T_e}=10,000 \pm 3,000$ K). This uncertainty has the largest affect on the output thermal flux -- adding an uncertainty of $\sim 14\%$ to the final predicted thermal fluxes. All mentioned sources of uncertainty and uncertainties in the empirical SFR relations are combined and accounted for using standard error propagation techniques. 

We would also like to note that the methodology of applying SFR relations to estimate thermal radio contribution does not rely on an IMF assumption, even though that is the case for the individual SFR relations. Since Equation 11 from \cite{murphyetal11} and our Equation \ref{SFRequation} were derived using the same ionizing photon rate from the same assumed IMF, variations in the ionizing photon rate (from variations in assumed IMF) balance each other out, when inverting Equation 11 from \citet{murphyetal11} to arrive at our Equation \ref{radioequation}. 

We can assess the effects of the various systematic uncertainties by noting the differences we see in the derived non-thermal spectral index maps between our methods. From difference maps in the non-thermal spectral index, we see a difference that is typically $\sim5\%$ in the central disk and $\sim10\%$  in the outer disk. Thus we conclude that resulting the non-thermal science is largely unaffected by systematic uncertainties in our methods.  


\section{Thermal Prediction Results}

\subsection{NGC 891}

\begin{figure*}
\centering
\includegraphics[scale=0.55]{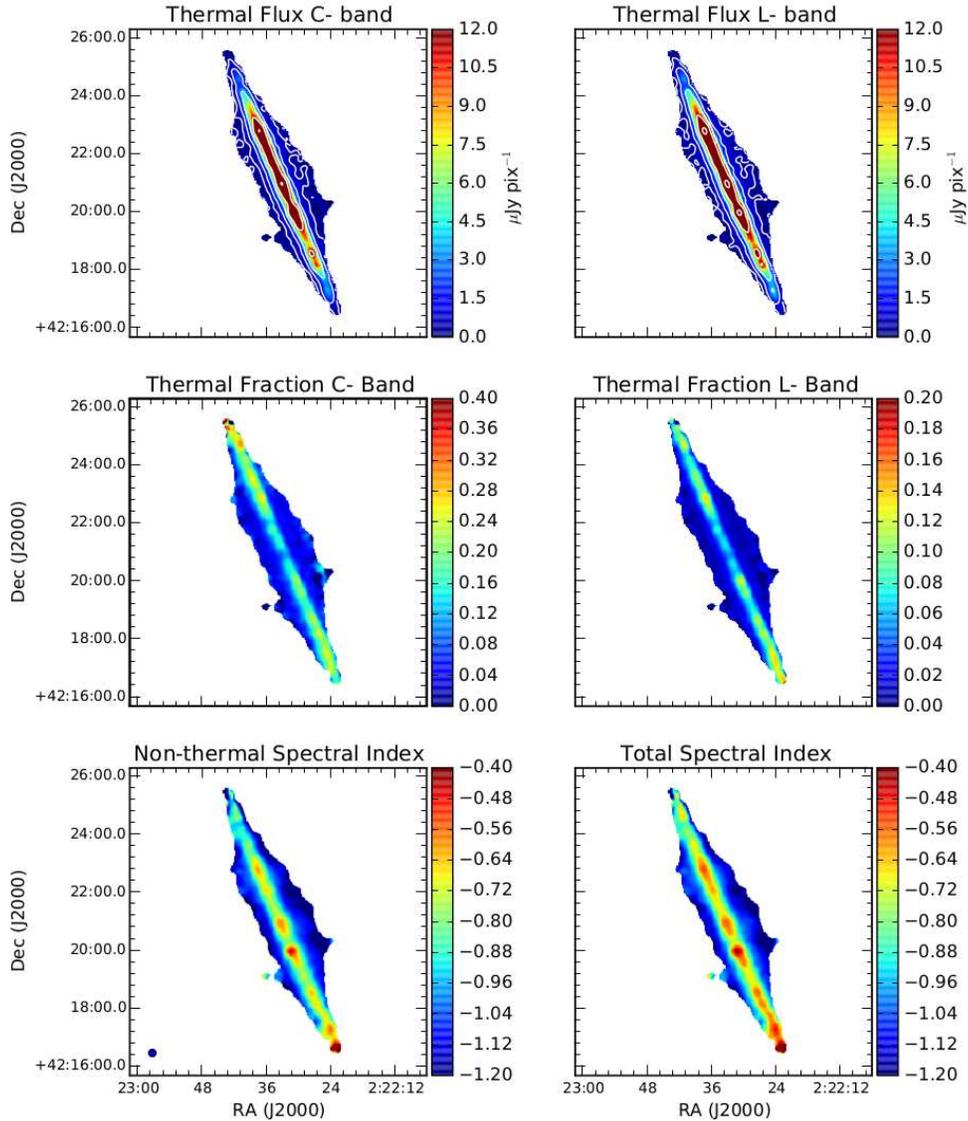}
\caption{\small Thermal Prediction results for NGC 891 using the $22$ micron Only method. The top row of panels show the predicted thermal flux in colorscale and contours. Contours correspond to ${\rm{0.45~\mu Jy~\rm{pix}^{-1}}\cdot (1,3, 9, 27...)}$. The colorscale and contours are at the same levels in both top panels. The central row of panels are the thermal fraction maps in C- and L-band using this method. The bottom row contain the non-thermal spectral index obtained from subtracting the predicted thermal component in C- and L-band, and the total spectral index distribution. All spectral indices shown are taken between C- and L-band. The total spectral index is obtained from observed radio maps before subtracting thermal emission. All panels are at $15\arcsec$ resolution, and the beam is shown in the lower left region of the lower left panel.}
\label{891jarmurphy}
\end{figure*}

\begin{figure*}
\centering
\includegraphics[scale=0.55]{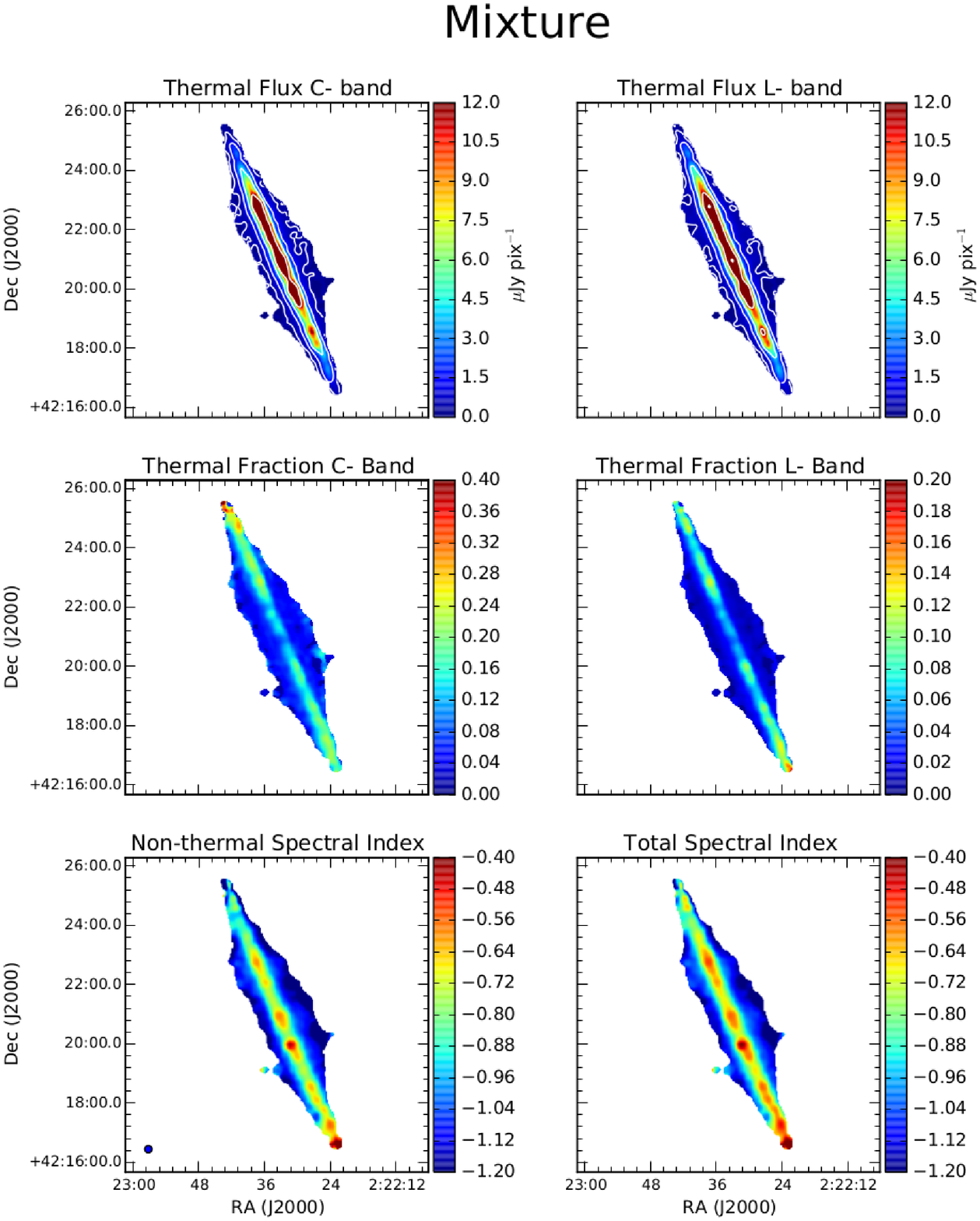}
\caption{\small Thermal Prediction results for NGC 891 using the mixture method. The panel layout, resolution, and contours are the same as Figure \ref{891jarmurphy}}
\label{891mixmurphy}
\end{figure*}

The thermal prediction results for NGC 891 using the $22$ micron only method and mixture method are shown in Figures \ref{891jarmurphy} and \ref{891mixmurphy}, respectively. Both methods show similar results. Within the disk, C- band thermal fractions range between $8\%-30\%$ in both methods. Outside of the disk, thermal fractions sit at $\sim 5\%$ in both methods, but the mixture method produces slightly more uniformly distributed extraplanar thermal emission. Also in both methods, we see evidence that the non-thermal spectral index steepens with vertical distance from the disk. 

Small differences between the two methods in each galaxy of the test sample can be seen by comparing regions along the plane of the galaxy within $\sim 3-5$ kpc of the center and outside of $\sim 3-5$ kpc, which we refer to as the inner and outer disks, respectively. In the case of NGC 891, both methods show a lower thermal fraction in the central disk ($\sim 11\%$ on average at C-band) than in the outer disk ($\sim22\%$ on average at C-band). For the mixture method, it is likely the H$\alpha$ encounters so much extinction that the calibrations used are no longer valid in the central disk. In fact, we see that the $22$ micron emission dominates the mixture in the central disk, due to H$\alpha$ extinction. The central deficiency of thermal emission in the $22$ micron only method could perhaps be due to extinction in the $22$ micron band, itself. NGC 891 is an almost perfectly edge-on disk and is a very dusty, late-type spiral, making it the most likely place for extinction at $22$ micron to occur. In principle, it is also possible that the central deficiency in thermal emission is truly due to decreased star formation in the central disk, and that CR diffusion produces non-thermal emission there. However, this scenario is likely unrealistic.

We note that the ratio [{\NII}]/H$\alpha$ was found to increase to $\sim 1.5$ at large heights above the disk ($z\sim 3$ kpc) of NGC 891 by \cite{rand98}. We explored the effects of this change in assumed [{\NII}]/H$\alpha$ by multiplying the input H$\alpha$ image for NGC 891 in the mixture method by a factor of $0.5$. We find that the while the thermal fractions above the disk increase by a factor of $\sim 1.2$, the non-thermal spectral index above the disk is almost completely unaffected, changing by a factor of only $1.005$ at most. Since the non-thermal results are almost completely unaffected, and the true [{\NII}] throughput in the H$\alpha$ filter would have to be assumed, we do not add a correction for [{\NII}]/H$\alpha$ variations with height above the disk.

\subsection{NGC 3044}

\begin{figure*}
\centering
\includegraphics[scale=0.50]{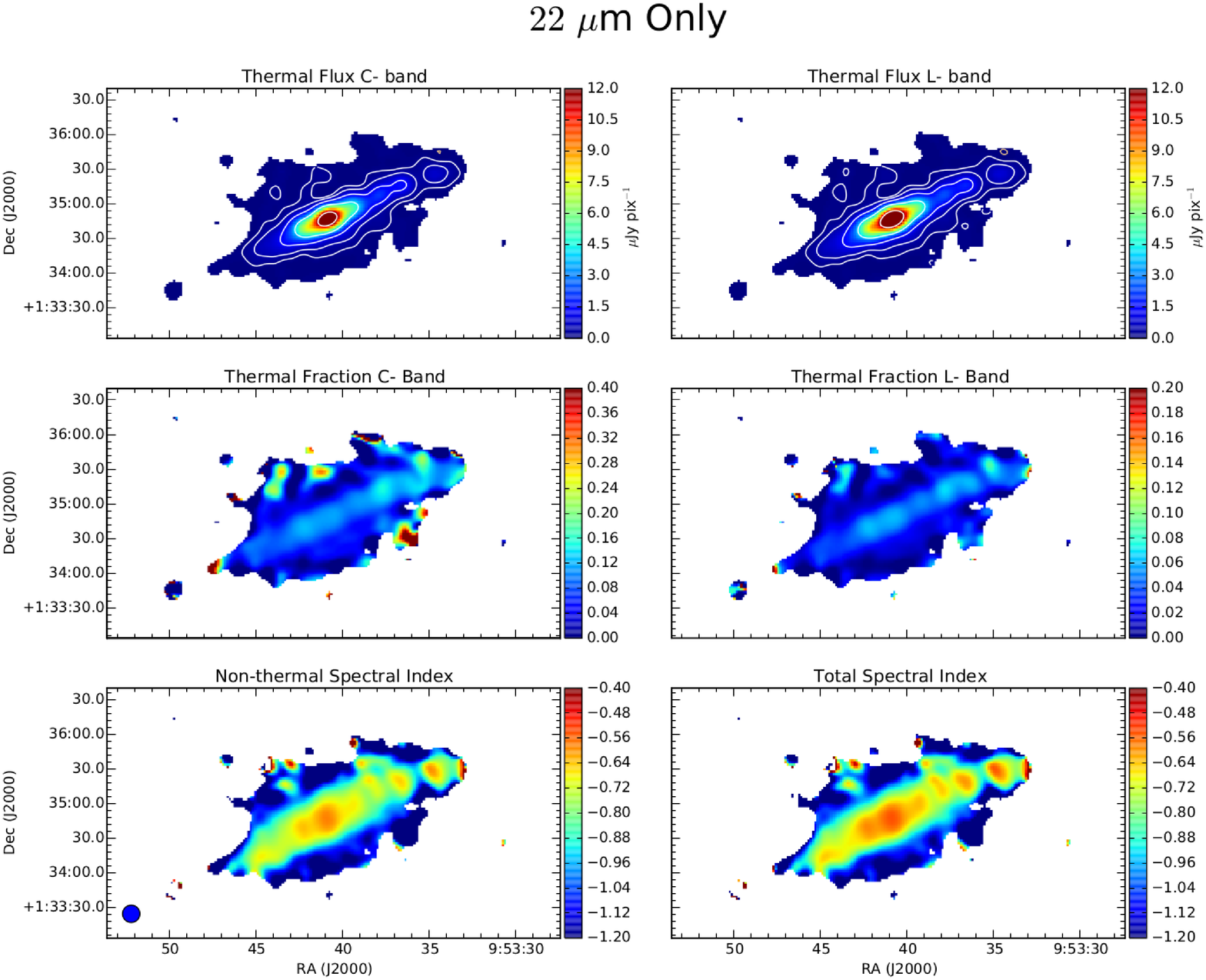}
\caption{\small Thermal Prediction results for NGC 3044 using the $22$ micron Only method. The top row of panels display the predicted thermal flux in colorscale and contours. Contours correspond to ${\rm{0.15\mu Jy~\rm{pix}^{-1}}\cdot (1,3, 9, 27...)}$. The colorscale and contours are at the same levels in both top panels. The central row of panels are the thermal fraction maps in C- and L-band using this method. The bottom row contain the non-thermal spectral index obtained from subtracting the predicted thermal component in C- and L-band, and the total spectral index distribution.  All spectral indices shown are taken between C- and L-band. The total spectral index is obtained from observed radio maps before subtracting thermal emission. All panels are at $15\arcsec$ resolution, and the beam is shown in the lower left region of the lower left panel.}
\label{3044jarmurphy}
\end{figure*}

\begin{figure*}
\centering
\includegraphics[scale=0.50]{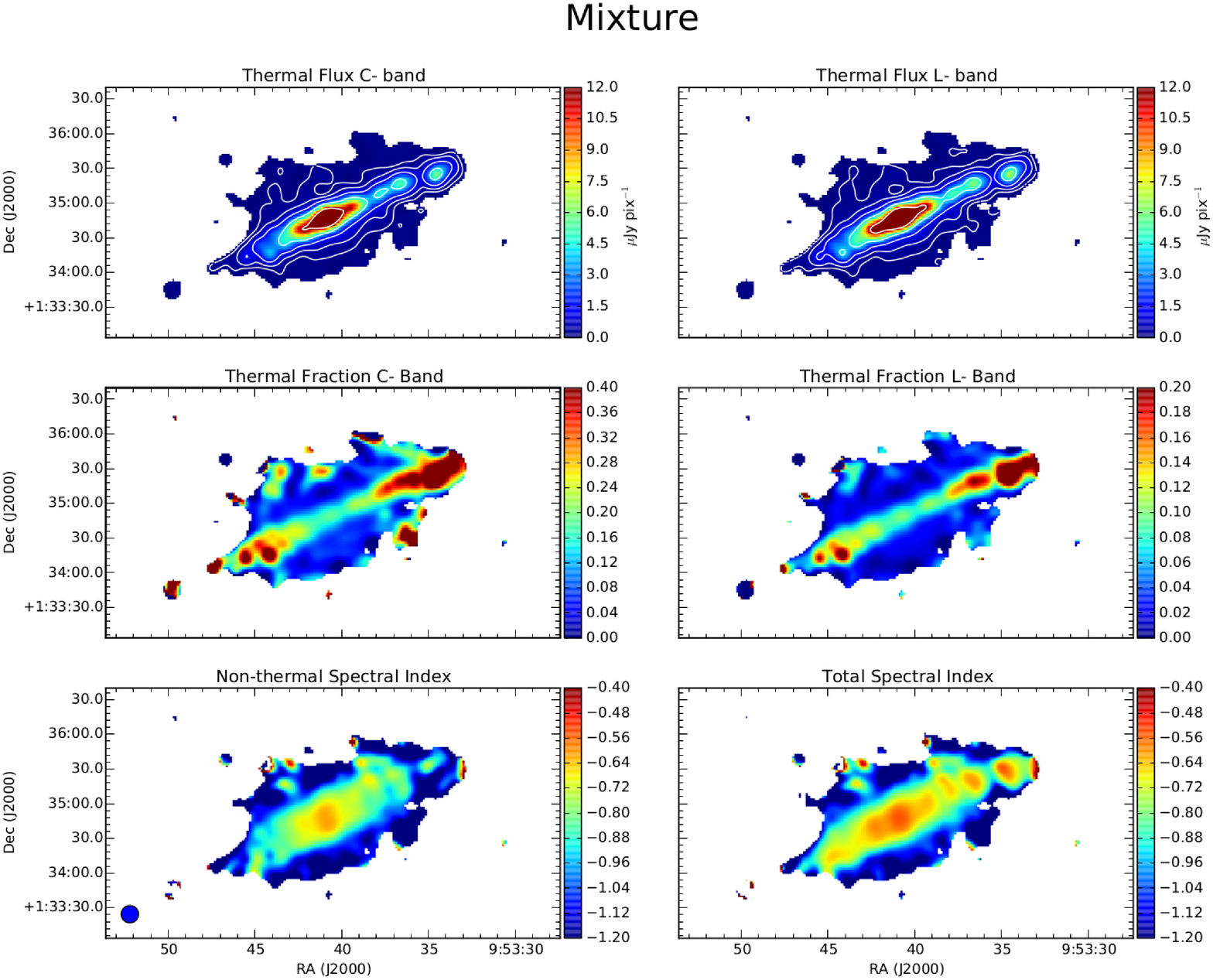}
\caption{\small Thermal Prediction results for NGC 3044 using the mixture method. The panel layout and contours are the same as Figure \ref{3044jarmurphy}}
\label{3044mixmurphy}
\end{figure*}

The thermal prediction results for NGC 3044 using the $22$ micron only method and mixture method are shown in Figures \ref{3044jarmurphy} and \ref{3044mixmurphy}, respectively. Unlike in NGC 891, the thermal morphology is different between the two methods. The $22$ micron only results show roughly constant thermal fraction values throughout the disk ranging between $\sim 10\% - 18\%$ at C-band. The mixture method shows larger thermal fractions in the outer disk than in the central disk, reaching as high as $80\%$ at C- band. These large thermal fraction regions trace bright {\HII} regions in the H$\alpha$ image. The H$\alpha$ emission is the most direct tracer of thermal emission, so the method that we adopt must reproduce at least the thermal emission implied by the H$\alpha$ emission. For this reason, the mixture method is more valid here than the $22$ micron only method, which is under-predicting the thermal contribution in the outer disk regions of this galaxy.

Vertically, we see that the mixture method shows more extraplanar thermal emission. As in NGC 891, we also see evidence for a steepening of the non-thermal spectral index with vertical distance from the disk. 

NGC 3044 is not as dusty or edge-on as NGC 891. Thus, the fact that we do not see a deficiency of thermal fraction in the central disk region of NGC 3044 using the $22$ micron only method implies that the $22$ micron emission in NGC 891 may be suffering from extinction. 

\subsection{NGC 4631}

\begin{figure*}
\centering
\includegraphics[scale=0.50]{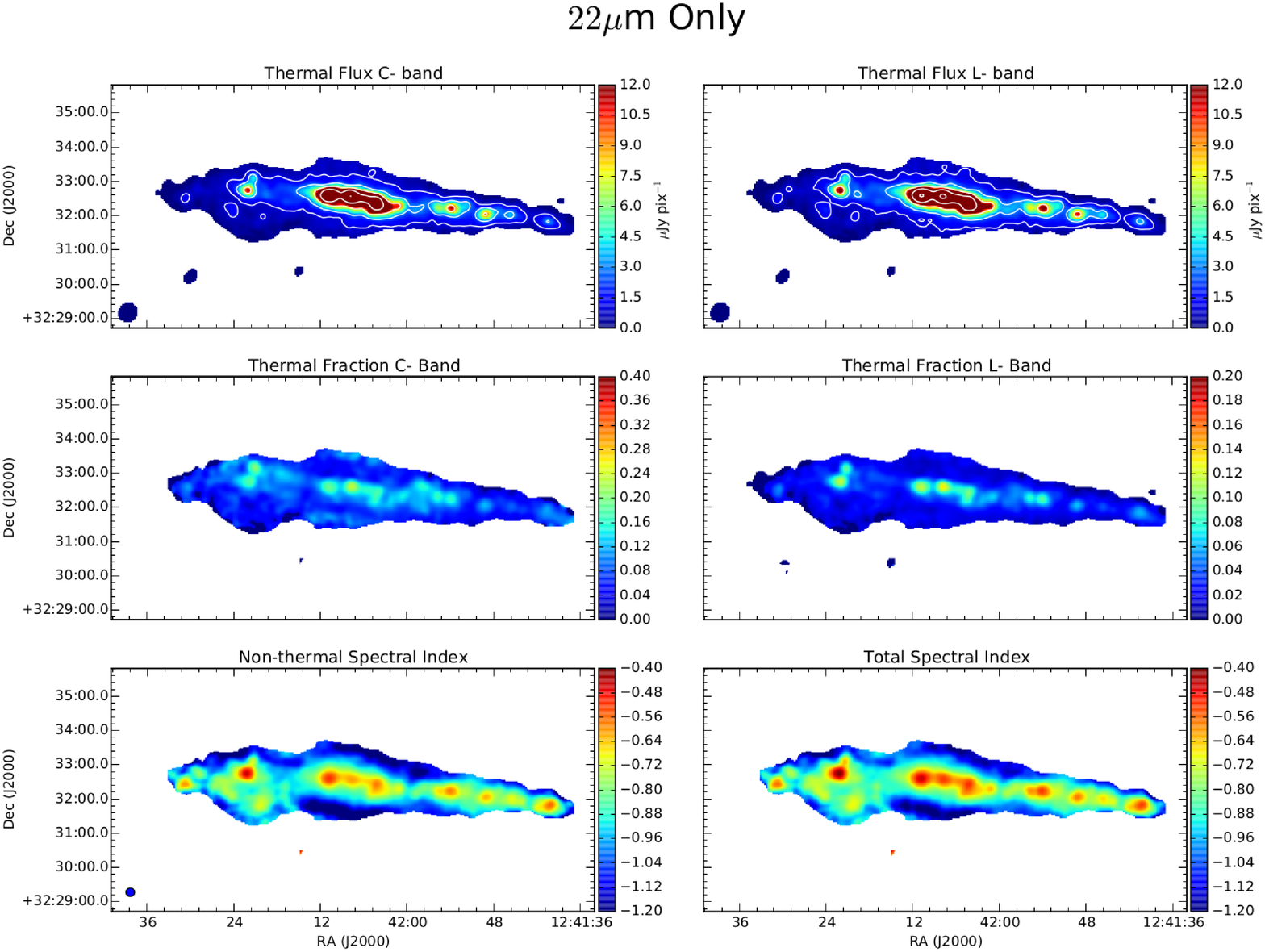}
\caption{\small Thermal Prediction results for NGC 4631 using the $22$ micron Only method. The top row of panels are the predicted thermal flux. Contours correspond to ${\rm{1\mu Jy~\rm{pix}^{-1}}\cdot (1,3, 9, 27...)}$. The colorscale and contours are at the same levels in both top panels. The central row of panels are the thermal fraction maps in C- and L-band using this method. The bottom row contain the non-thermal spectral index obtained from subtracting the predicted thermal component in C- and L-band, and the total spectral index distribution. All spectral indices shown are taken between C- and L-band. The total spectral index is obtained from observed radio maps before subtracting thermal emission. All panels are at $15\arcsec$ resolution, and the beam is shown in the lower left region of the lower left panel.}
\label{4631jarmurphy}
\end{figure*}

\begin{figure*}
\centering
\includegraphics[scale=0.50]{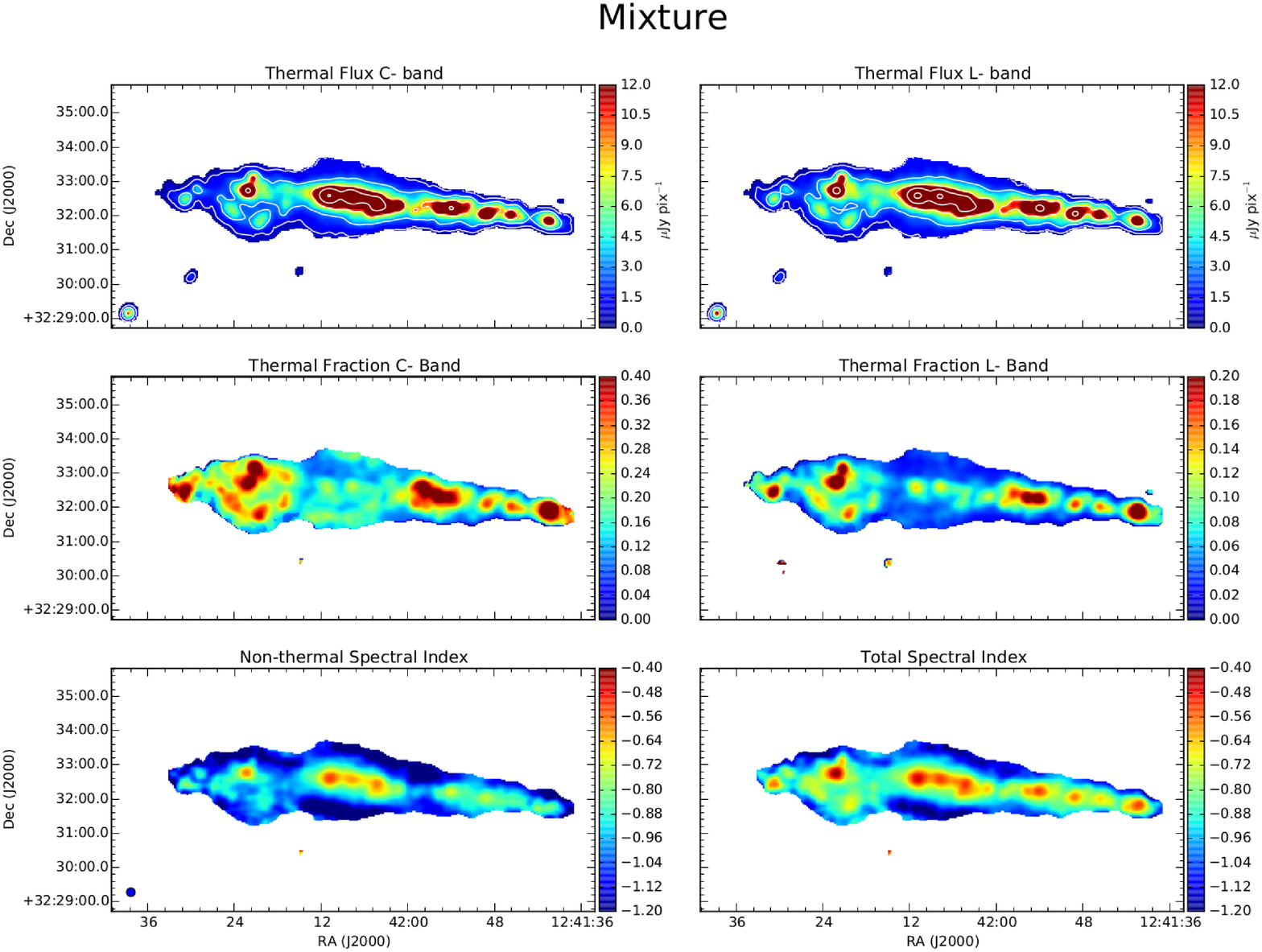}
\caption{\small Thermal Prediction results for NGC 4631 using the mixture method. The panel layout, resolution, and contours are the same as Figure \ref{4631jarmurphy}}
\label{4631mixmurphy}
\end{figure*}

The thermal prediction results for NGC 4631 using the $22$ micron only method and mixture method are shown in Figures \ref{4631jarmurphy} and \ref{4631mixmurphy}, respectively. The case of NGC 4631 is similar to NGC 3044; we find a roughly uniform thermal fraction distribution throughout the disk using the $22$ micron only method, while the outer disk shows considerably more thermal emission using the mixture method. From the $22$ micron only method, disk thermal fraction values range from $5\% - 13\%$ (minimum in L-band, maximum in C-band). For the mixture method, disk thermal fraction values range from $7\% - 60\%$. The mixture method also produces more extraplanar thermal emission than the $22$ micron only method.  

As is the case for NGC 3044, bright {\HII} regions in the outer disk drive the predicted thermal emission up in the outer disk. The $22$ micron only method does not reproduce the intensity of thermal emission in those regions that is implied by the H$\alpha$ component in the mixture, and thus leads to an under-prediction. We also note that the central disk predicted thermal emission values agree within $5\%$ for both methods in NGC 4631. 

We list the integrated thermal prediction results for all three galaxies in Table \ref{inttable1}.

\begin{table*}
\center
\begin{tabular}{ c c c c c c  } 

 \hline
 \hline

Galaxy & L$_{H\alpha}$ ($10^{41}$ erg s$^{-1}$) & F$_{22\mu\mathrm{m}}$  (Jy) & F$_{\mathrm{C-band}}$ (mJy) &   F$_{\rm{C-band}}^{T}$   Mixture (mJy) & F$_{\rm{C-band}}^{T}$ $22$ micron Only (mJy)
 \\
 \hline 
 NGC 891 & 0.16 $\pm$ 0.027 & 5.84 $\pm$ 0.18 & 205.86 $\pm$ 6.9 &  24.6 $\pm$ 5.4 & 26.3 $\pm$ 4.2 \\ 
 NGC 3044 & 1.04 $\pm$ 0.25 & 0.7 $\pm$ 0.03 & 37.5 $\pm$ 0.9 & 6.2 $\pm$ 1.2 & 3.7 $\pm$ 0.5  \\ 
 NGC 4631 & 2.02 $\pm$ 0.33 &  7.88 $\pm$ 0.24 & 284.4 $\pm 7.4$ & 59.3 $\pm$ 12.8 & 34.1 $\pm$ 5.5 \\
 \hline
 \hline
\end{tabular}
\caption{Integrated results of various calibrations and methods. Column values from left to right: galaxy name, integrated H$\alpha$ luminosity, integrated $22$ micron flux density, integrated total C-band flux density, integrated predicted thermal C-band flux density via the mixture method, integrated predicted thermal C-band flux density via the $22$ micron only method. Thermal prediction results are provided for C- band only, as L- band results are all scaled versions of the C- band results.}
\label{inttable1}

\end{table*}


\subsection{H$\alpha$ and $22$ Micron Contribution to the Mixture Method}

\begin{figure*}
\centering
\includegraphics[scale=0.36]{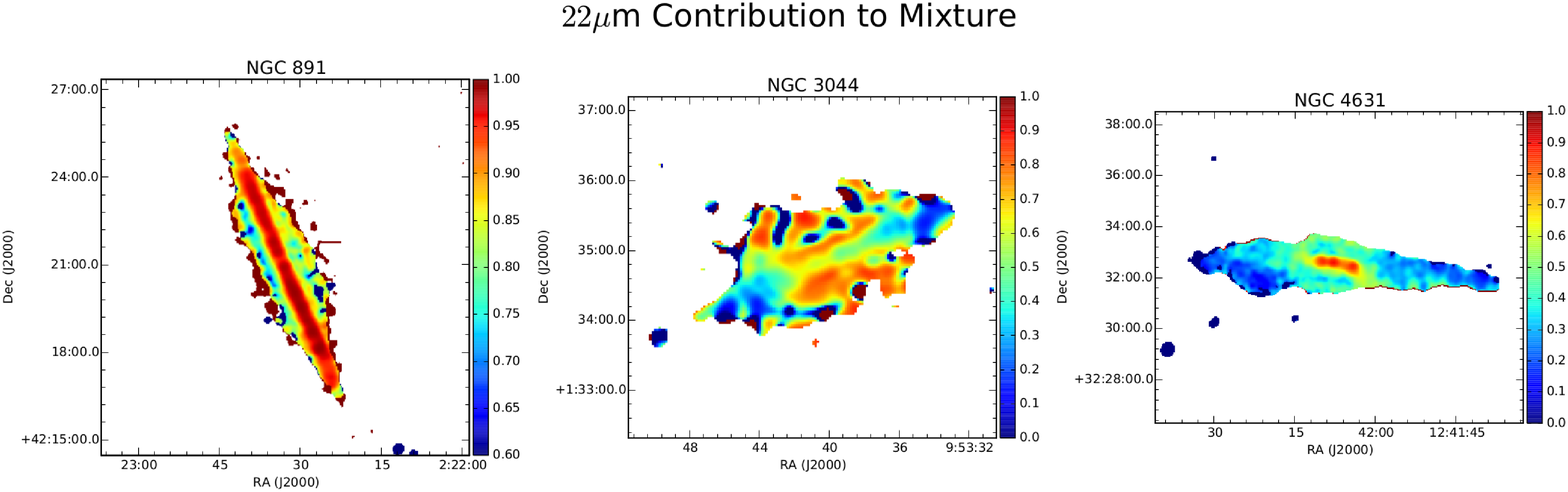}
\caption{\small Maps of the fractional contribution of $22$ micron emission to the total mixture method results in each of the three test sample galaxies. Large values represent regions where the $22$ micron emission is the dominant component to the mixture method results.}
\label{contribution}
\end{figure*}

It is useful to analyze the contribution of each component to the mixture method, to better understand the morphological behavior of this method's thermal prediction results. We create maps of the $22$ micron image's contribution to the mixture. Specifically, we show maps of $\frac{\rm{a}\cdot \nu \rm{L}_\nu (24\mu\rm{m})} { \rm{L}(\rm{H}\alpha_{\rm{obs}})+\rm{a}\cdot\nu \rm{L}_\nu (24\mu\rm{m})}$. Thus, the magnitudes in the maps are the fractional contribution of the $22$ micron image's contribution to the total mixture method. These maps are shown in Figure \ref{contribution}.

The results contrast from galaxy to galaxy. In NGC 891, we see that the $22$ micron emission is almost completely dominant in the entirety of the disk, while the H$\alpha$ becomes slightly more present above and below the disk. This effect could be related to the dust lanes at high vertical distance found in \citet{howketal97}. In NGC 3044 we find that the H$\alpha$ seems to be more dominant along the disk, with larger contribution from the $22$ micron imaging in bands above the central disk. Since NGC 3044 is a disturbed disk seen slightly less than edge-on, it is possible that this effect is due to emission from the front-side and back-side of the disk masquerading as vertical emission. We also see clear H$\alpha$ dominance in {\HII} region clumps in the disk. For NGC 4631, we see that the H$\alpha$ is largely dominant throughout, except in the central disk. Extreme values that arise at the outermost edges of the emission are not considered reliable, and are likely due to wings in the WISE PSF (see \citealt{cutrietal12}).

\section{Infrared Extinction}

In NGC 891, we have identified possible signs of extinction in the WISE $22$ micron data. These signs manifest as a central depression in thermal fraction using both the mixture and $22$ micron only methods. To explore the existence of this extinction and its implications to the non-thermal emission, we employ ancillary Spitzer MIPS $70$ micron and $160$ micron imaging, and analyze the FIR total flux properties of a larger sample from the Infrared Astronomical Satellite (IRAS) Revised Bright Galaxy Sample (RBGS; \citealt{sandersetal03}).

\subsection{Spitzer MIPS Imaging}  

\begin{figure*}
\centering
\includegraphics[scale=0.45]{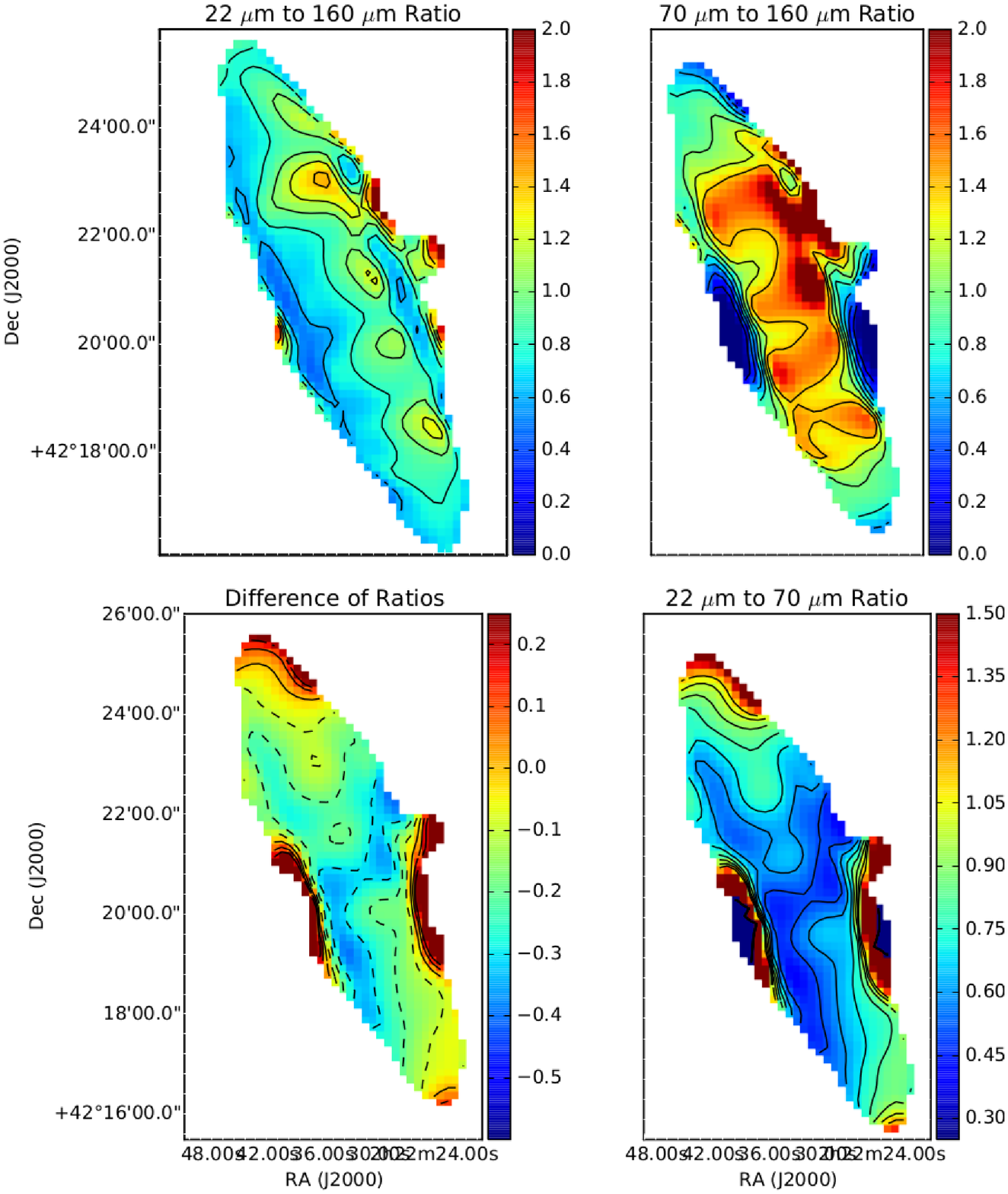}
\caption{\small Infrared ratio maps for NGC 891. Values near the edges of emission are likely unreliable, due to background uncertainties and low signal-to-noise.}
\label{891mipsratios}
\end{figure*}

\begin{figure*}
\centering
\includegraphics[scale=0.35]{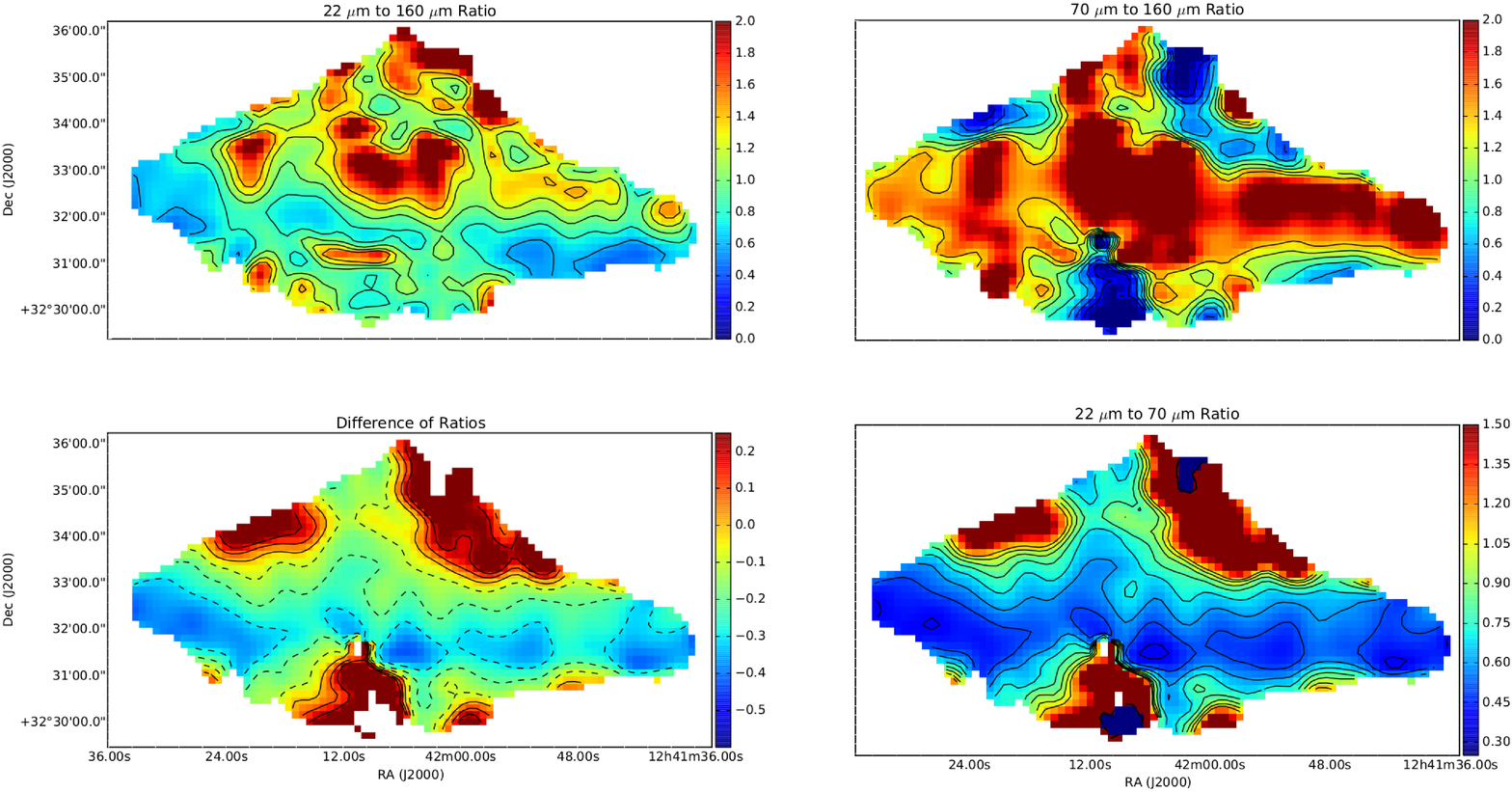}
\caption{\small Infrared ratio maps for NGC 4631. Values near the edges of emission are likely unreliable, due to background uncertainties and low signal-to-noise.}
\label{4631mipsratios}
\end{figure*}

We analyze existing \textit{Spitzer} MIPS $70$ micron and $160$ micron imaging for NGC 891 \citep{bendoetal12} and NGC 4631 \citep{daleetal09}. No MIPS data exists for NGC 3044. We plot the ratios of MIPS band maps and WISE $22$ micron imaging, all convolved to Gaussian beams using the kernels from \citet{anianoetal11}, and smoothed to match the resolution of the $160$ micron image in Figure \ref{891mipsratios}. In the plot of $22/70$ micron ratio (lower right panel), the ratio appears to be lowest in the central region, and increases radially. This could potentially point to extinction in the central disk, as also motivated by the previous thermal prediction results. The same map ratios are plotted for NGC 4631 in Figure \ref{4631mipsratios}. The $22/70$ micron ratio does not appear to vary radially, as in NGC 891. 

We note that the $70$ and $100$ micron emission likely originates from larger dust grains than the grains responsible for $22$ micron emission. These larger grains do not necessarily trace the population of the smaller grains, which could explain the differing morphologies. 
We also note the possible contribution of cirrus to the $70$ micron imaging, which would arise from dust heating by sources other than star formation. The existence of cirrus in the $70$ micron imaging could lead to an over-estimate of the thermal contribution in some regions.

We use the MIPS $70$ micron image to independently predict the thermal radio component via the SFR calibration from \citet{calzettietal10}:

\begin{equation}
\left( \frac{\rm{SFR}_{70\mu\rm{m}}}{\rm{M_{\odot}}{yr}^{-1}} \right) = \frac{\rm{L}_{\nu} (70~\mu \rm{m})}{1.7\times10^{43} ~\rm{erg} \cdot \rm{s}^{-1}}
\end{equation}

\begin{figure*}
\centering
\includegraphics[scale=0.55]{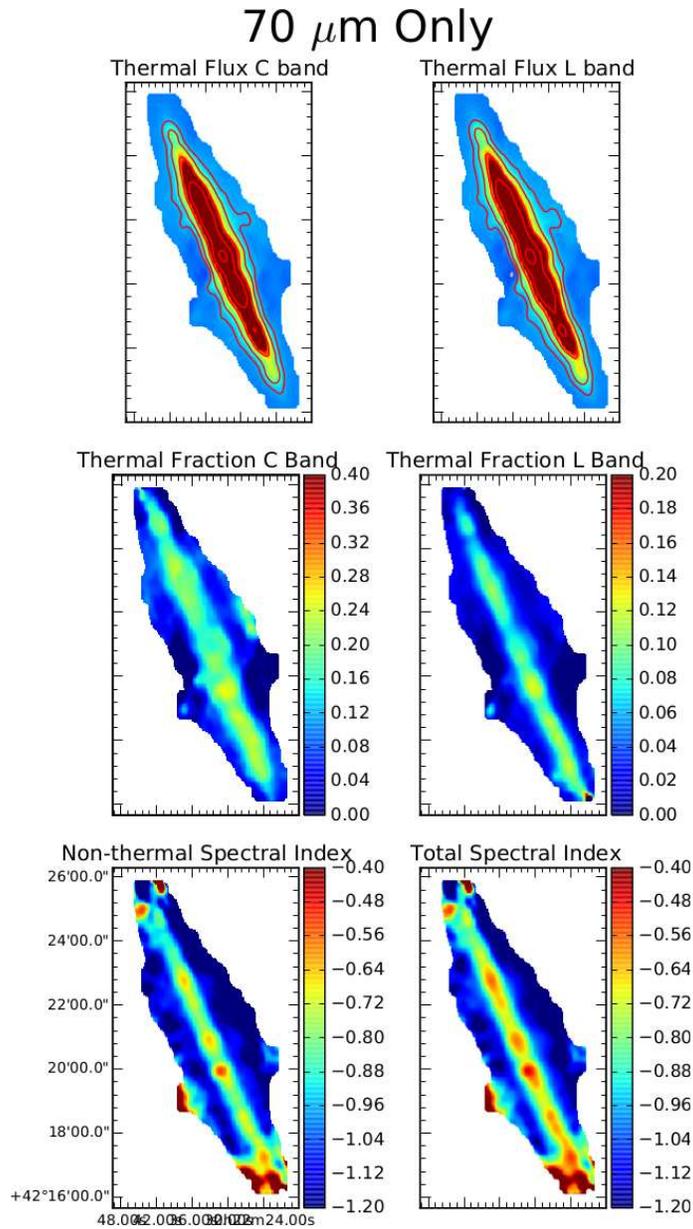}
\caption{\small Thermal prediction results using $70$ micron MIPS data for NGC 891. The resolution is $20\arcsec$ in each panel. The panel layout and contours are the same as in Figure \ref{891jarmurphy}.}
\label{891_70pred}
\end{figure*}

From this step, we use Equation \ref{radioequation} to predict the corresponding thermal radio component in L- and C- bands. The results of the $70$ micron only thermal prediction are shown in Figure \ref{891_70pred} for NGC 891. The roughly constant thermal fraction along the entirety of the disk of NGC 891, as seen in the $70$ micron only prediction, does seem to be consistent with our speculation that the $22$ micron emission is suffering from extinction in the central disk.  

\subsection{Electron Temperature}

We next consider the possibility that variations in the electron temperature could account for the thermal fraction being observed to be lower in the central disk regions of galaxies. Through varying the assumed electron temperature between $7000$ K and $13,000$ K (the expected range of $\rm{T_e}$ in galaxy disks), we find that the thermal fraction would vary from $11\%$ to $21\%$ in the central disk region of NGC 891, for the same input data. Thus, variations in $\rm{T_e}$ of $~\sim3000$ K  could explain the variations in thermal fraction along the disk that we see in both methods. In fact, $\rm{T_e}$ is likely to decrease in regions of high metallicity, such as the central disk region of NGC 891, which would lower the thermal fraction of the radio emission. For example, \citet{zuritaetal12} measured electron temperatures as low as $\sim7100$ K in regions at low galactocentric distance and high metallicity in M31. With this in mind, and the fact that the $70$ micron and $160$ micron emission are likely from different dust populations than the $22$ micron emission, we deem the presence of extinction in the central disk regions of our galaxies' $22$ micron data inconclusive to this point. 

\subsection{Independent Evidence for Mid-IR Extinction}

Here we explore further the issue of extinction at 22 microns in edge-on galaxies (and arrive at a global estimate of the effect) via a different direction. First, we note two lines of evidence that highlight the concern.
The first comes from studies of the mid-infrared extinction law in the disk of the Milky Way.
Measurements of the extinction at wavelength $\lambda$ relative to that in the K$_{\rm{s}}$ filter, A$_{\lambda}/$A$_{\rm{Ks}}$, for wavelengths of $22$ and $24$ micron have been made by \citet{flahertyetal07}, \citet{chapmanetal09}, and \citet{xueetal16}. The measured values span a wide range, with some authors finding evidence for higher values found on sightlines with lower A$_{\rm{Ks}}$. As an illustration of how significant extinction at these wavelengths might be in an edge-on disk, we consider the following illustrative example. We assume R$_{\rm{V}} \equiv$A$_{\rm{V}}/($A$_{\rm{B}}-$A$_{\rm{V}})=3.1$, N$_{\rm{H}}/$A$_{\rm{V}}=1.9 \times 10^{21}$ cm$^{-2}$ mag$^{-1}$ where N$_{\rm{H}}$ is the H column density \citep{bohlinetal78}, 
a representative gas
density of $1$ cm$^{-3}$, a path length of $10$ kpc, and A$_{24}/$A$_{\rm{Ks}} = 0.4$. For these assumptions, A$_{24}$ is significant, at $0.7$ mag.
Second, in \citet{lietal16}, 
it was reported that the average total IR (dominated by far IR emission) to mid-IR
luminosity ratio of the CHANG-ES sample is significantly higher than that in the samples of \citet{riekeetal09} and \citet{calzettietal10}, which contain star-forming galaxies with a range of inclinations.
Assuming negligible extinction at far-infrared wavelengths, a dependence of this ratio on inclination is
an indication of excess mid-IR extinction in edge-ons. However, the properties of the CHANG-ES sample
are not matched to those of the other two, so that biases may be present.

To optimally address the issue via such a ratio, a large sample of galaxies with a range of inclinations (including a significant number of edge-on galaxies) observed at mid-and far-infrared wavelengths in a uniform way is desirable. Unfortunately, no such samples exist for galaxies observed with any combination of Spitzer MIPS, Herschel, and WISE photometers. We therefore turn to the IRAS Revised Bright Galaxy Catalog (RBGS, \citealt{sandersetal03}), where fluxes are available for 629 galaxies. The RBGS is a flux-limited sample of all extragalactic objects observed by IRAS with 60 micron fluxes $> 5.4$ Jy.
Our main
focus is the variation of the $25$ to $100$ micron flux ratio, $f_{25}/f_{100}$. Ideally, one would want to
examine this ratio as a function of inclination, but we restrict ourselves to comparing only two
samples of galaxies, edge-on and “non-edge-on”, given the size of the survey.

\begin{figure}
\centering
\includegraphics[scale=0.65]{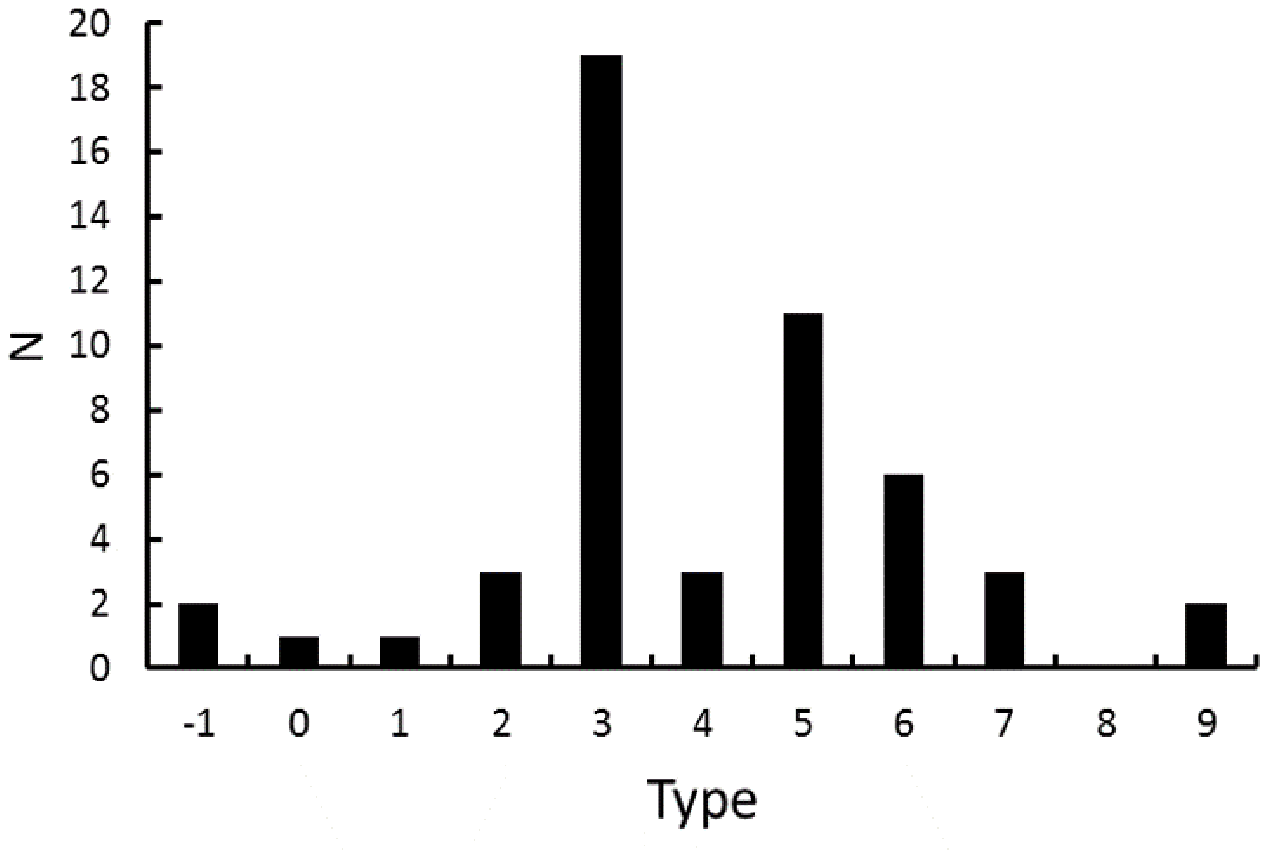}
\includegraphics[scale=0.65]{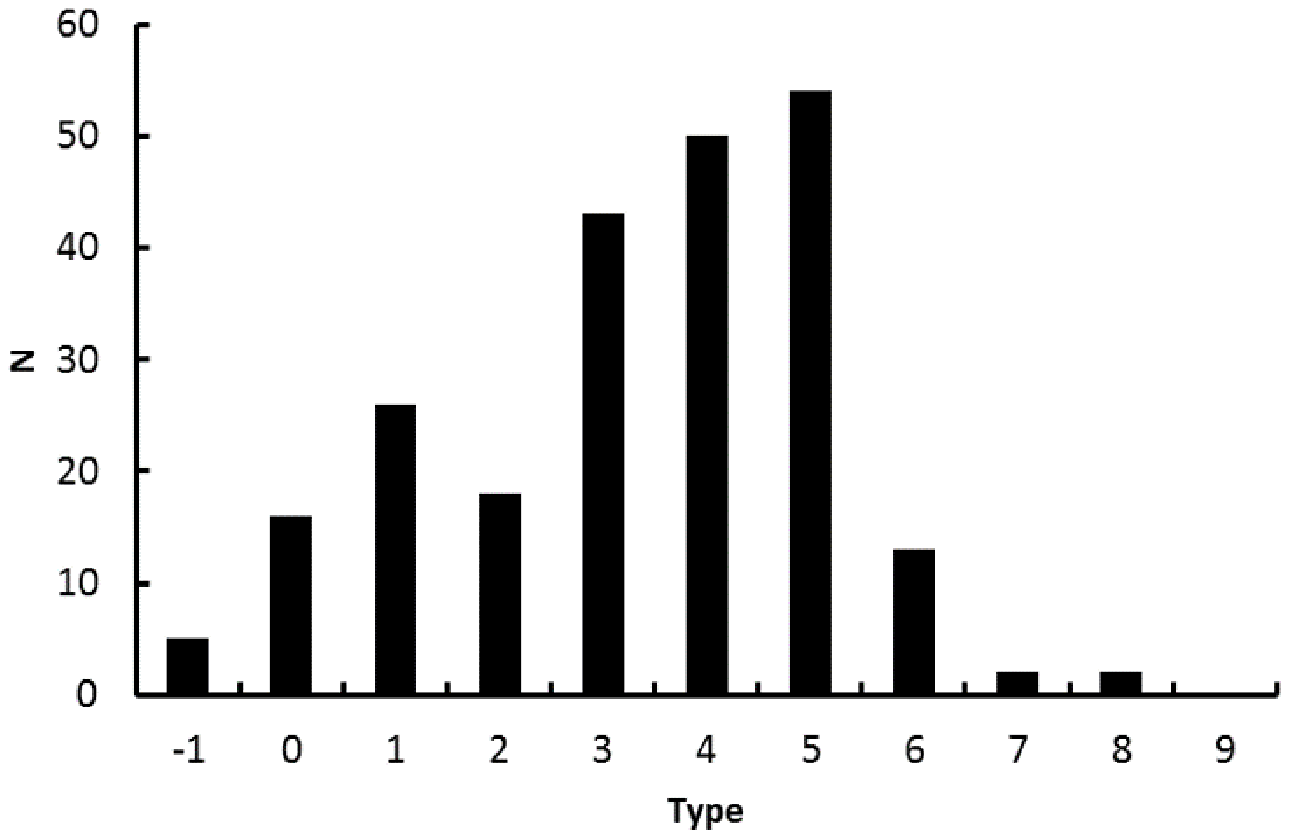}
\caption{\small Distribution of galaxies by type in the  edge-on (top) and non-edge-on (bottom) samples from the RBGS.}
\label{histtypes}
\end{figure}

We limited our study to disk galaxies (S0 to Sm) with no confusion flags on their fluxes. While $23$
CHANG-ES galaxies are in the RBGS, we increased the edge-on sample to $52$ by examining optical
images of all the RBGS galaxies to find ones that appeared to be very edge-on in that they showed no
sign of face-on disk structures (such as spiral structure) viewed at a high inclination. We then need a
non-edge-on comparison sample matched as closely as possible to the properties of the edge-on
sample. For this sample, we restricted axial ratios (from Lyon-Meudon Extragalactic Database, LEDA) to be $\leq 2$ to exclude very inclined galaxies. The non-edge-on sample has 227 galaxies. For disk galaxies in the full RBGS sample, we found that $f_{25}/f_{100}$ varies systematically with type and with L$_{\rm{TIR}}$ (the $8-1000$ IR luminosity \citep{perault87}, in units of L$_{\odot}$), with earlier type and higher L$_{\rm{TIR}}$ disk galaxies showing higher $f_{25}/f_{100}$ (Figure \ref{fratiovsLTIR}). We therefore restricted the non-edge-on sample to span the same range of type (-1 to 9, or SO$^+$ to Sm)  and L$_{\rm{TIR}}$ (log L$_{\rm{TIR}}=9.22$ to $11.32$) as the edge-on sample. However, the distribution of galaxy types in each sample
is somewhat different, with earlier type disks being somewhat more common in the edge-on sample (Figure \ref{histtypes}). 

Therefore, to control for the variation of $f_{25}/f_{100}$ with type, we calculated median
values of $f_{25}/f_{100}$ for the non-edge-ons and the edge-ons for four bins of type, namely -1 to 2, 3 to 5, 6
to 7, and 8 to 9. Within these ranges, the variation of $f_{25}/f_{100}$ in the full RBGS sample is relatively small. Although outliers only have a small effect on the medians, two outliers in the edge-on sample, M82 (a starburst) and NGC 4388, with $f_{25}/f_{100}$ ratios $>10\sigma$ from the median of the other galaxies in their bin, were removed. Two galaxies, NGC 1068 and NGC 4151  (both Seyferts), were removed from the non-edge-on sample via the same criterion. We then formed the ratio of these two medians for each of the four type bins, and then averaged these ratios for the bins, weighted by the total number of galaxies in both samples in each bin. The distribution of L$_{\rm{TIR}}$ values for the two samples is very similar, hence we
do not need to correct for any L$_{\rm{TIR}}$ bias. The non-edge-on sample extends to somewhat larger distances than the edge-on sample, but we found no correlation of $f_{25}/f_{100}$ with distance in the full RBGS that would suggest distance bias is a concern.

\begin{figure}
\centering
\includegraphics[scale=0.55]{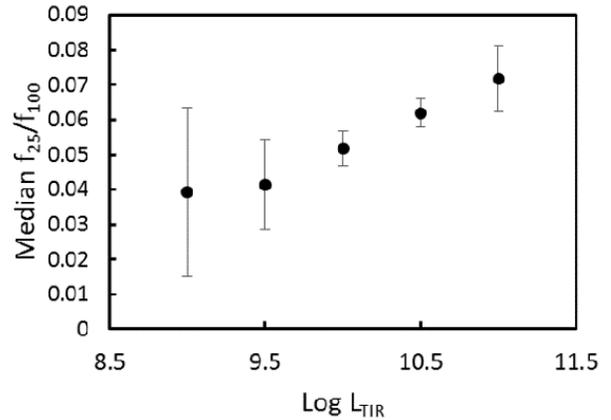}
\caption{\small Median values of $f_{25}/f_{100}$ vs. Log(L$_{\rm{TIR}}$) for disk galaxies in the full RBGS sample. }
\label{fratiovsLTIR}
\end{figure}

\begin{figure}
\centering
\includegraphics[scale=0.55]{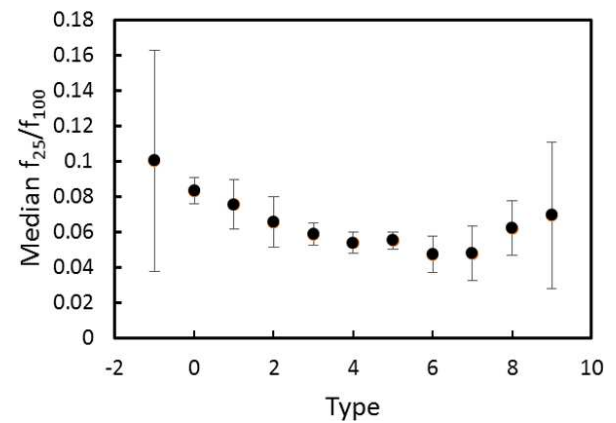}
\caption{\small Median values of $f_{25}/f_{100}$ vs. type for disk galaxies in the full RBGS sample. }
\label{fratiovsLTIR}
\end{figure}

\begin{figure}
\centering
\includegraphics[scale=0.65]{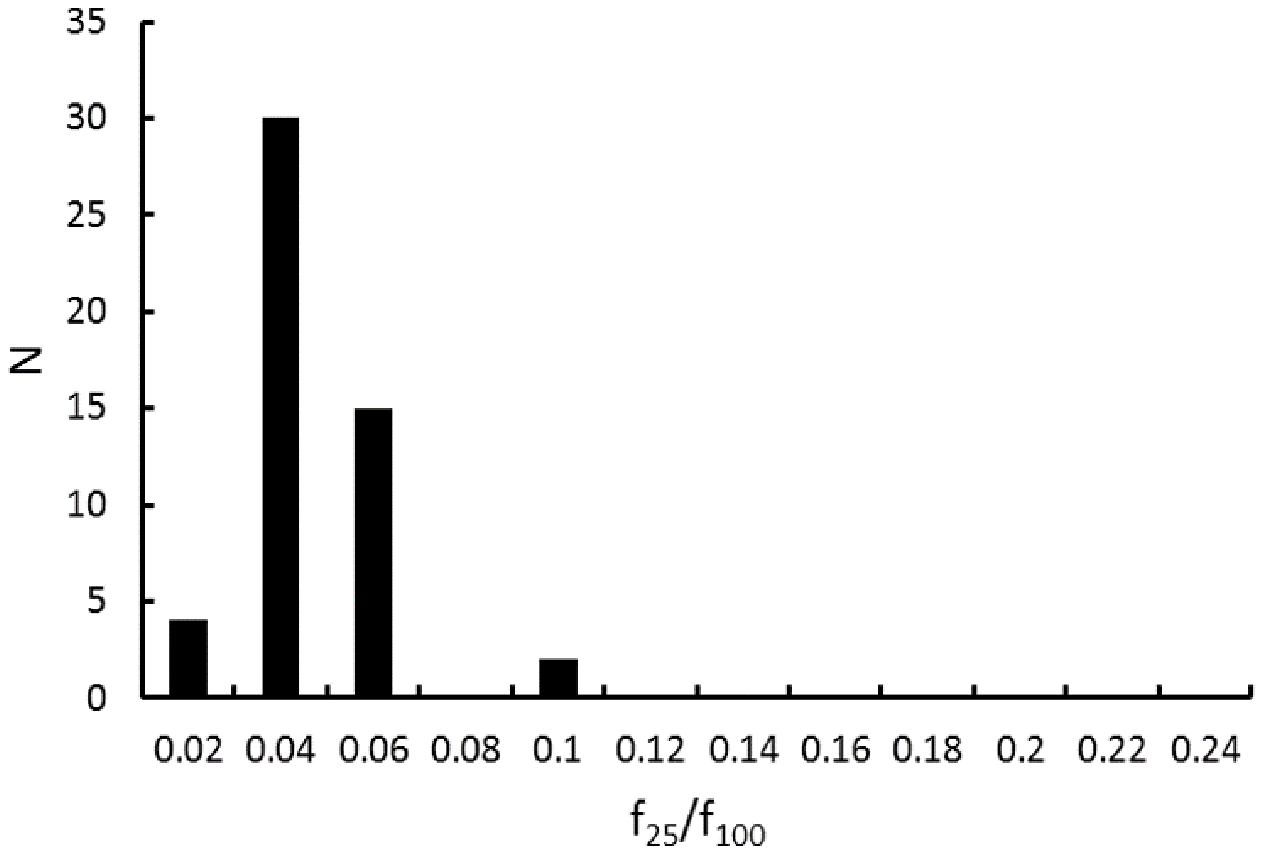}
\includegraphics[scale=0.65]{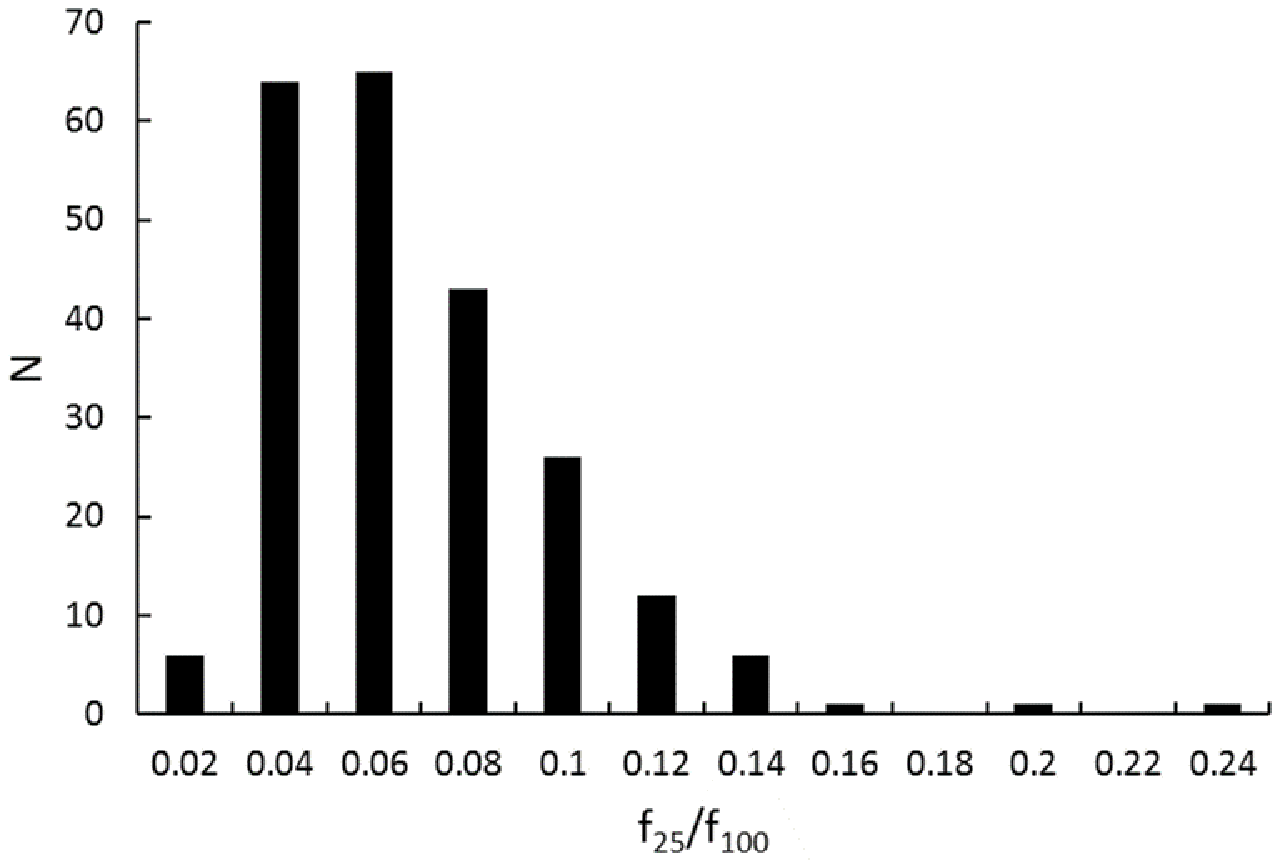}
\caption{\small Distribution of galaxies by $f_{25}/f_{100}$ in the edge-on (top) and non-edge-on (bottom) samples from the RBGS.}
\label{histfratio}
\end{figure}

Histograms of $f_{25}/f_{100}$ for the two samples are shown in Figure \ref{histfratio}. They appear to be different, with
a high $f_{25}/f_{100}$ tail in the non-edge-on sample not present in the edge-on sample. Controlling for type
as above, the resulting value for the $f_{25}/f_{100}$ contrast between non-edge-ons and edge-ons is $1.35 \pm 0.11$. This differs insignificantly from the value found without controlling for type, which is $1.36\pm0.09$. A Kolmogorov-Smirnov test was performed to test whether the values of $f_{25}/f_{100}$ for the two samples
could be drawn from the same parent population. The probability of this being the case is rejected with $>99.9$\% confidence.

We therefore conclude, assuming negligible extinction at 100 microns, that edge-on galaxies have an
extra attenuation of 1.36 in their IRAS 25 micron emission relative to galaxies with axial ratios $\leq 2$. We assume that the same value applies in the WISE $22$ micron band, although a more thorough analysis would account for the relative spectral response of the $22$ and $25$ micron bands and the (uncertain) shape of the measured extinction law between $22$ and $25$ microns (see below), but that is beyond the scope of the current paper. 

Although our main interest is $f_{25}/f_{100}$, we repeat the analysis for the other IRAS bands. For $f_{12}/f_{100}$
and $f_{60}/f_{100}$, much weaker trends with type were found in the full RBGS. We therefore calculate
median values for each sample, and find that $f_{12}/f_{100}$ and $f_{60}/f_{100}$ are higher for non-edge-ons by
$1.12 \pm 0.12$ and $1.13 \pm 0.13$, respectively - thus not significant contrasts. The rough wavelength
range of the 12 micron IRAS response was about 8 to 15 microns (\citealt{iras}; IRAS Explanatory Supplement),
which, as dust models (e.g. \citealt{weingartneretal01}) indicate, includes the silicate absorption
feature centered at 9.7 microns as well as a low extinction trough (although less pronounced in the ice
mantle model of \citealt{wangetal15}) at around 14 microns. These models would suggest that the
lower 12 micron relative to 25 micron extinction implied by our results is surprising. However, there
may be marginal support for this result from the Milky Way observations. With the aforementioned
caveat, we note that values of A$_{24}/$A$_{\rm{Ks}}$ 
from \citet{chapmanetal09} toward low A$_{\rm{Ks}}$ sightlines are
somewhat higher than values of A$_{12}/$A$_{\rm{Ks}}$ (\citealt{lutz99} and \citealt{xueetal16}), although A$_{22}/$A$_{\rm{Ks}}$ and A$_{24}/$A$_{\rm{Ks}}$ found by \citet{xueetal16} indicate the opposite trend.

\section{Recommended Thermal Prediction Method}

The RBGS results suggest that the WISE $22$ micron emission is not entirely optically thin for all edge-on galaxies. The RBGS analysis brought to light a factor of 1.36 of extinction, which represents the average extinction across entire galaxies. A single, consistent method of thermal prediction is needed for edge-on galaxies. In all galaxy cases, the mixture method captures thermal flux that must necessarily be present as implied by the H$\alpha$ contribution, in the vertical direction. In NGC 3044 and NGC 4631, we see evidence of this in the outer disk, as well. With this in mind, we elect to use the mixture method over the $22$ micron only method for edge-on galaxies.

Additionally, we add a correction for extinction to the $22$ micron data. This correction will be applied via an adjustment to the $a$ factor in the mixture method corresponding to the average extinction value determined by the RBGS analysis above. Accounting for the factor of $1.36$ average extinction within the original $a$ value of $0.031$ makes the new $a$ value increase to $0.042$. While correct on average, this correction does not account for potential variations in extinction on a galaxy-to-galaxy basis, which would require additional gas and dust information to establish. The mixture method relation valid for edge-on and extremely dusty galaxies is included in Equation \ref{mix_edgeon}.

\begin{equation}
\label{mix_edgeon}
\rm{L(H\alpha_{corr}}) = L(\rm{H\alpha_{obs}}) + 0.042 \cdot \nu L_{\nu}(24 \rm{\mu m})
\end{equation}

\begin{figure*}
\centering
\includegraphics[scale=0.53]{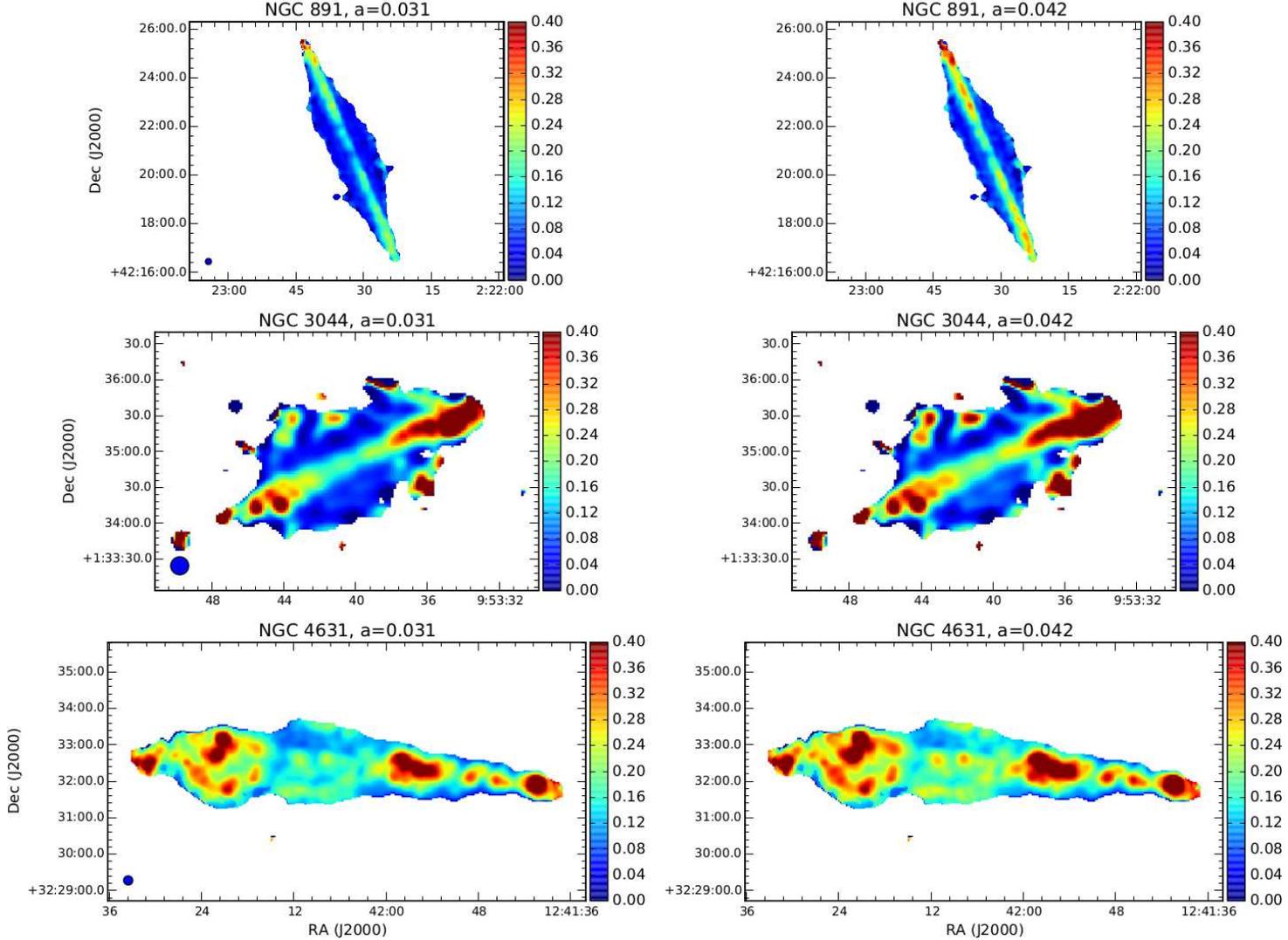}
\caption{\small Comparison of C-band thermal fraction results using $a=0.031$ (left column) and $a=0.042$ (right column). All panels are shown at $15\arcsec$ resolution, and the beam is shown in the lower left region of the panels in the left column.}
\label{thermalfractionsplot}
\end{figure*}

We show the integrated thermal prediction results with the recommended method in Table \ref{inttable2}, and a comparison of the thermal fraction results in C-band between $a=0.031$ and $a=0.042$ in Figure \ref{thermalfractionsplot}. The results of the mixture method with $a=0.042$ show a slight increase in the overall thermal fraction as compared to those of $a=0.031$, as expected. 

\begin{table*}
\center
\begin{tabular}{ c c c c c c c c} 

 \hline
 \hline

Galaxy & F$_{\mathrm{C-band}}$ (mJy) & F$_{\mathrm{L-band}}$ (mJy) & F$_{\rm{C-band}}^{T}$ (mJy) & F$_{\rm{L-band}}^{T}$ (mJy) & $\%^{T}_{\rm{C-band}}$ & $\%^{T}_{\rm{L-band}}$ & $\Delta \%^{T}_{\rm{C-band}}$ \\
 \hline 
 NGC 891 & 205.86 $\pm$ 6.9 & 743.9 $\pm$ 14.9  & 32.0 $\pm$ 7.0 & 36.8 $\pm$ 8.1 & 15.5 $\pm$ 4.1 & 4.9 $\pm$ 1.3 & $\sim12$ \\ 
 NGC 3044 & 37.5 $\pm$ 0.9 & 104.2 $\pm$ 2.1   & 7.3 $\pm$ 1.4 & 8.4 $\pm$ 1.6 & 19.5 $\pm$ 4.3 & 8.1 $\pm$ 1.7 & $\sim23$ \\ 
 NGC 4631 & 284.4 $\pm$ 7.4 & 1083.0 $\pm$ 37.0 & 69.8 $\pm$ 15.1 & 80.2 $\pm$ 17.3 & 24.5 $\pm$ 6.1 & 7.4 $\pm$ 1.9 & $\sim 14$ \\
 \hline
 \hline
\end{tabular}
\caption{Integrated results of the mixture method with correction for $22$ micron extinction. F$_{\mathrm{C-band}}$ and F$_{\mathrm{L-band}}$ are the total integrated C-band and L-band flux densities from the CHANG-ES data. F$_{\rm{C-band}}^{T}$ and F$_{\rm{L-band}}^{T}$ are the integrated predicted thermal flux densities at C-band and L-band, respectively. The percentage thermal fraction values at C-band and L-band are included, under  $\%^{T}_{\rm{C-band}}$ and $\%^{T}_{\rm{L-band}}$. In the rightmost column, we include $\Delta \%^{T}_{\rm{C-band}}$, the estimated mean change in C-band thermal fraction from the inner disk to the outer disk in the $22$ micron extinction corrected method. }
\label{inttable2}

\end{table*}

One could envision a method of thermal prediction where the $22$ micron data is corrected for extinction by varying the $a$ factor in the mixture method to represent changes in the path length along different lines of sight. We do not currently have the gas and dust properties as would be mapped by further ancillary data on hand for each galaxy in our sample. Thus, a path-length-dependent method would not be trivial, and would require further assumptions. 

Additionally, a more complex treatment of this `a' factor could take into account its variation with dust column density and height above the plane, e.g. by introducing a dependence of `a' on $\nu \rm{L}_{\nu}(24\mu m)$, hence replacing $a \cdot \nu \rm{L}_{\nu}(24\mu m)$ by a nonlinear term $a \cdot \nu \rm{L}_{\nu}(24\mu m)^{\rm{x}}$ on the right-hand side of Equation 2. Constraining this `x' exponent would require further information on the expected thermal contribution, a priori. The best extinction-free observational tool for obtaining the expected thermal contribution in edge on galaxies would likely be radio mapping at frequencies where the thermal component dominates ($\sim30$ GHz).  

\section{Non-thermal Spectral Index Behavior}
\label{nthspixsection}

\begin{figure*}
\centering
\includegraphics[scale=0.53]{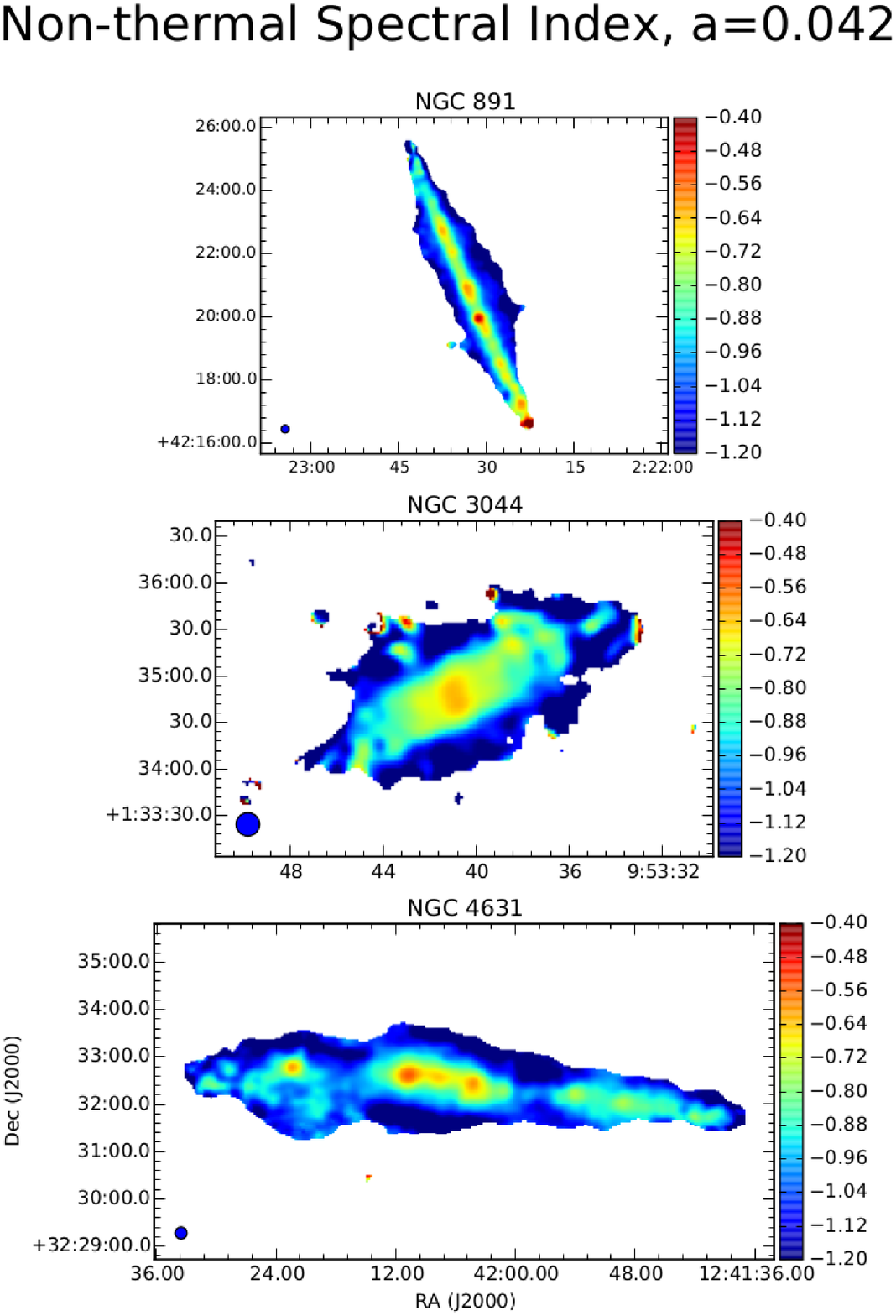}
\caption{\small Non-thermal spectral index results using the recommended method of thermal prediction. All panels are shown at $15\arcsec$ resolution, and the beam is shown in the lower left region of each panel.}
\label{nthspixplot}
\end{figure*}

Maps of the non-thermal spectral index for the three test sample galaxies using the final recommended method is included in Figure \ref{nthspixplot}. Vertical variations non-thermal spectral index distribution can be seen in all galaxies, independent of the method used to obtain the thermal prediction. We analyze the change in non-thermal spectral index with vertical distance from the major axis by defining five rectangular boxes oriented along the minor axis for each galaxy. The boxes have a height of 1 kpc, and are contiguously oriented to span $\pm 2$ kpc from the major axis, with the central box encompassing the inner $\pm 0.5$ kpc. The lengths of the boxes were chosen so that each box contains as much emission as possible. We calculate the mean non-thermal spectral index value in each box. The results of this analysis are included in Table \ref{nthspixtable}. 

\begin{table}[h!]
\center
\begin{tabular}{ c c c c  } 

 \hline
 \hline

Region & NGC 891 & NGC 3044 & NGC 4631 \\
 \hline
 $+2$ kpc & -1.16 $\pm$ 0.28 & -0.82 $\pm$ 0.17 & -1.18 $\pm$ 0.28 \\
 $+1$ kpc & -0.98 $\pm$ 0.24 & -0.75 $\pm$ 0.16 &  -0.98 $\pm$ 0.24\\
 $0$ kpc &  -0.73 $\pm$ 0.18 & -0.75 $\pm$ 0.16 & -0.72 $\pm$ 0.17 \\
 $-1$ kpc & -1.02 $\pm$ 0.25 & -0.77 $\pm$ 0.16 &-0.95 $\pm$ 0.23\\
 $-2$ kpc & -1.17 $\pm$ 0.28 & -0.89 $\pm$ 0.19 & -1.24 $\pm$ 0.30 \\
 \hline
 \hline
\end{tabular}
\caption{Mean values of non-thermal spectral index from rectangular boxes oriented along the major axis at different heights above the major axis. The box layout is described in Section \ref{nthspixsection}. Uncertainties from both the thermal predictions and radio data are considered in the quoted uncertainties. }
\label{nthspixtable}

\end{table}

The results in Table \ref{nthspixtable} show clear evidence for a steepening of the non-thermal spectral index with increased distance from the major axis, on average. We note this analysis does not account for radial variations in non-thermal spectral index. However, radial variations in non-thermal spectral index can be clearly seen in the morphology included in Figure \ref{nthspixplot}.  These tend to be closely associated with regions of high SFR. We note one interesting exception: the furthest southwestern extention of the disk in NGC 891 shows a knot of very flat non-thermal spectral index. Normally, this is indicative of star formation processes. However, in this case, there is very little thermal flux in that region.

A previous study of the total spectral index of NGC 891 by \citet{hummeletal91} shows a relatively flat disk spectral index ($\alpha\sim -0.4$) that steepens with vertical distance to $\alpha\sim-0.9$ in the halo. A similar behavior is seen in NGC 4631 by \citet{hummeletal90} with data at 327 MHz and 1.49 GHz, where $\alpha \sim -0.45$ in the disk, and approaches $\alpha \sim -1.0$ in the halo. \citet{kleinetal84} estimated the global non-thermal spectral index in NGC 891 to be $\alpha\sim -0.9$, which is similar to our results for that galaxy. Our new sensitive data, coupled with secure thermal subtraction methodology, now leads to high-quality determination of the vertical variation of the non-thermal spectral index and therefore opens the possibility for careful cosmic ray propagation modeling.

Studies of the non-thermal radio continuum component that do not take the morphology of thermal emission into account (e.g. \citealt{heesenetal09}, \citealt{mulcahyetal14}) may suffer from some inaccuracies. In general, if magnetic field strengths are calculated with the assumption of energy equipartition between the magnetic field and cosmic rays, the field strength will increase with steepening of the non-thermal spectral index. In this study, we find that typical non-thermal spectral indices are flattest in the disk. Thus, previous studies that assume a singular and steep non-thermal spectral index throughout the galaxy may be under-estimating the magnetic field strength.  
Since synchrotron losses scale with magnetic field strength, underestimating the magnetic field strength will lead to an underestimate of synchrotron losses. The opposite is true in halo regions, where the non-thermal spectral index may be steeper than that assumed.

\section{Summary and Conclusions}

We have analyzed various methods of star formation rate estimation, and thus thermal radio continuum prediction, in a test sample of three edge-on galaxies (NGC 891, NGC 3044, and NGC 4631) included in the CHANG-ES sample. The goal of the analysis is to determine a method for producing consistent and accurate estimates of the thermal radio component in edge-on galaxies.

We have used H$\alpha$ imaging from various sources, and \textit{WISE} $22$ micron maps from \cite{jarrettetal12} with enhanced resolution to create SFR maps for the test sample using two methods: a mixture of H$\alpha$ and $22$ micron, and $22$ micron only. We find that the $22$ micron only method misses thermal flux that the H$\alpha$ data is sensitive to, both in the outer disk and extraplanar regions of the test sample galaxies. We also find possible evidence for extinction in the $22$ micron band itself for NGC 891.

We have further explored the possibility of $22$ micron extinction in NGC 891 by utilizing \textit{Spitzer} MIPS $70$ micron and $160$ micron imaging of both NGC 891 and NGC 4631. We find that flux ratio maps of $22/70$ micron show slight evidence for extinction at $22$ micron in the central disk of NGC 891 only. Also, a thermal prediction map created from $70$ micron data only corroborates this. 

We have explored the possibility that the potential extinction features in the thermal prediction results can be explained through variations in electron temperature. We find that the difference in observed thermal fraction from the central disk to outer disk can be explained by varying the electron temperature within its uncertainty. This would be consistent with a lowering of the electron temperature in the central disk with increased metallicity. Currently, this possibility cannot be ruled out, however a more detailed study of the physical conditions in edge-on galaxies is needed to more plausibly explain the behavior of our results. 

We have compiled an expanded sample of edge-on and face-on galaxies was compiled within the RBGS, which contains fluxes in all IRAS bands. Flux ratios of $25$ to $100$ micron show that the edge-on sample is characteristically lower in flux than the face-on sample, which points to extinction. This extinction acts as a lowering of the $25$ micron flux by a factor of $1.36$ on average for the edge-on sample. 

Using this independent measurement of extinction, we apply a correction factor of $1.36$ to the $22$ micron data in the mixture method (Equation \ref{Halpha+24microns}), which corresponds to raising the IR weighting factor in the mixture method to a$=0.042$. 

We recommend that edge-on galaxies have their thermal radio component estimated using the mixture method. Additionally for edge-ons, we recommend a correction of the $22$ micron data for extinction by increasing the $a$ factor in the mixture method to $0.042$. 

We outline evidence that the non-thermal spectral index steepens with vertical distance for the galaxies of the test sample in Section \ref{nthspixsection}. This vertical steepening is likely due to CR aging; CR injection sites, such as supernova remnants and young O-B stars tend to be located within galaxy disks. As CRs propagate upward out of the disk, the higher energy CRs lose energy. This energy loss is observed as a steepening of the spectral index in regions with an older population of CRs. We also see evidence for radial variations in non-thermal spectral index, which are closely associated with star formation regions within the galaxy disk, indicating CR injection occurs in complexes of high star formation. There is an exception seen in the southwestern extension of the disk in NGC 891. That region shows a complex of extremely flat non-thermal spectral index, with very minimal thermal flux associated with the region. However, more information is needed on the region to fully understand what is driving the peculiar behavior. 

\acknowledgements 

We would like to thank Fatemeh Tabatabaei for useful conversations on this work. We would also like to thank the anonymous referee for feedback that improved this paper. This material is based upon work supported by the National Science Foundation Graduate Research Fellowship under Grant No. 127229 to CJV. This material is also based on work partially supported by the National Science Foundation under Grant Nos. AST-0908126, AST-1615594, and AST-1616513 to RAMW and RJR. This research has made use of the VizieR catalogue access tool, CDS, Strasbourg, France. The National Radio Astronomy Observatory is a facility of the National Science Foundation operated under cooperative agreement by Associated Universities, Inc. 

\newpage

\newpage
\bibliography{refs_thermal.bib}
\bibliographystyle{apj}

\end{document}